\RequirePackage{fix-cm}
\documentclass[epjc3]{svjour3}  
\smartqed  
\RequirePackage{graphicx}
\RequirePackage{amssymb}
\RequirePackage{textcomp}
\RequirePackage{cite}
\RequirePackage{xcolor}
\RequirePackage{amsmath}
\RequirePackage{subfig}

\def\eq#1{(\ref{#1})}
\journalname{Eur. Phys. J. C}
\begin{document}

\title{Gravitational axial perturbations and quasinormal modes of loop quantum black holes
}


\author{M.B.Cruz\thanksref{e1,addr1}
        \and
        C. A. S. Silva \thanksref{e2,addr2} 
				\and
				F.A.Brito \thanksref{e3,addr1, addr3}
}

\thankstext{e1}{messiasdebritocruz@gmail.com}
\thankstext{e2}{carlos.silva@ifpb.edu.br}
\thankstext{e3}{fabrito@df.ufcg.edu.br}


\institute{Departamento de F\'{i}sica, Universidade Federal da Para\'{i}ba,
58051-970, Jo\~{a}o Pessoa, Para\'{i}ba, Brazil. \label{addr1}
           \and
Instituto Federal de Educa\c{c}\~{a}o Ci\^{e}ncia e Tecnologia da Para\'{i}ba (IFPB),\\ Campus Campina Grande - Rua Tranquilino Coelho Lemos, 
671, Jardim Dinam\'{e}rica
I. \label{addr2}
           \and
Departamento de F\'{i}sica, Universidade Federal de Campina Grande
Caixa Postal 10071, 58429-900 Campina Grande, Para\'{i}ba, Brazil.
           \label{addr3} \label{addr3}
}

\date{Received: date / Accepted: date}

\maketitle

\begin{abstract}
Loop Quantum Gravity (LQG) is a theory that proposes a way 
to model the behavior of the spacetime in situations where its atomic characteristic arises. Among these situations,
the spacetime behavior near the Big Bang or black hole's singularity.
The detection of gravitational waves, on the other hand, has opened the way to new perspectives in the investigation of the spacetime structure. 
In this work, by the use of a WKB method introduced by Schutz and Will \cite{Schutz:1985zz}, and after improved by Iyer and Will \cite{s.iyer-prd35},
we study the gravitational wave spectrum emitted by loop quantum black holes, which correspond to a quantized version of the Schwarzschild
spacetime by LQG techniques. From the results obtained, loop quantum black holes have been shown stable under
axial gravitational perturbations.

\end{abstract}

\section{Introduction}
\label{intro}

One of most exciting predictions of general relativity is the existence of black holes, 
objects from which no physical bodies or signals can get loose of their drag due to its strong gravitational field. 
Going far beyond astrophysics, black holes appear as objects that may help us to clarify
one of the most intriguing bone of contention in the current days, the quantum
nature of gravity. It is because, in the presence of a black hole's strong gravitational field, quantum features of spacetime must be manifested \cite{Mathur:2005zp, Nozari:2008gp, Silva:2008kh, Silva:2010ir, Fazeli:2010zz, Kim:2011fh, Silva:2014jda}.

Loop quantum gravity, on the other hand, is a theory that
has given ascent to models that provide
a portrait of the quantum features of spacetime unveiled by a black hole.
In particular, in the context of this theory, a black hole metric known as
Loop Quantum Black Hole (LQBH), or self-dual black hole, has been proposed \cite{Modesto:2008im, Modesto:2009ve}.
This solution corresponds to a quantum corrected Schwarzschild solution and possess the interesting
property of self-duality. From this property, the black hole 
singularity is removed and replaced by
another asymptotically flat region, which is an expected effect in a quantum gravity regime. 
Moreover, LQBHs have been pointed as a possible candidate for dark matter \cite{Modesto:2009ve, Aragao:2016vuy} and as the building blocks of a holographic version of loop
quantum cosmology \cite{Silva:2015qna}.

In order to move black holes 
from a simple mathematical solution of the gravitational equations
to objects whose existence in nature is possible,
a key point consists in to investigate black hole's stability under perturbations.
It is due to the fact that an isolated black hole would never be found in nature. 
In fact, complex distributions of matter such as accretion disks, galactic nuclei, strong magnetic
fields, other stars, etc are always present around black holes, which in turn actively interact with 
their surroundings.
Even if all macroscopic objects and
fields in space have been removed, a black hole will interact with the vacuum
around it, creating pairs of particles and evaporating due
to Hawking phenomena.
Besides, in the first moments after
its formation, a black hole is in a perturbed state due to gravitational collapse of matter.
In this way, a real black hole will be always in a perturbed state.

A black hole's response to a perturbation occurs by emitting gravitational waves whose evolution
corresponds, firstly, to a 
relatively short period of initial outburst of radiation followed by a phase where 
the black hole get going to vibrating into exponentially decaying oscillations, ``quasinormal modes'', whose
frequencies and decay times depend only on the intrinsic characteristics of the black hole itself, 
being indifferent to the
details of the collapse. 
Finally, at a very large time, the quasinormal modes are slapped down by power-law or 
exponential late-time tails.


The issue of black hole stability under perturbations was firstly addressed by Regge and Wheeler \cite{Regge:1957td}, and by Zerilli \cite{Zerilli:1971wd}, 
which based on the black hole perturbation theory,
have demonstrated the stability of the Schwarzschild metric.
The methods used are familiar from
quantum mechanics: perturbations caused by an external (e.g., gravitational or electromagnetic) field are taken into 
account as waves scattering off the respective potential. 
It is due to the fact that 
the formalism provided by Regge-Wheller and Zerilli removes the angular dependence in the perturbation variables 
by the use of a tensorial generalization of
the spherical harmonics which makes possible to translate the solution of the perturbed Einstein's equations in the form of a 
Schrodinger-like wave equation treatment. 
Posteriorly, the Regge-Wheeler/Zerilli formalism was extended to the charged \cite{Zerilli:1974ai, Moncrief:1974ng, Moncrief:1974gw} 
and rotating \cite{Teukolsky:1972my, Teukolsky:1974yv}
black hole scenarios.  A full description of the black hole perturbation framework can be found in the text
by Chandrasekhar \cite{s.chandrasekhar-mtbh}.

The most valuable phase in the evolution of black hole perturbations is given
by its quasinormal modes which can give us information not only about the black hole stability, but also, as emphasized by Berti in \cite{Berti:2004md}, 
"`how much stable it is"'. In other words, quasinormal modes tell us which timescale a black hole radiate away its matter contend after formation.
By the way, in this context the prefix "`quasi"' means
that the black hole consists in an open system that loses energy due to the emission of gravitational waves. 
The issue of black hole quasinormal modes is interesting not only by the investigation of
black hole stability, but also because gravitational waves have been pointed
as a possible experimental way to make contact with black holes and, in this way, with the quantum spacetime
characteristics revealed by them. Indeed, recent results from LIGO, which has detected
a gravitational wave signal from black holes \cite{Abbott:2016blz, TheLIGOScientific:2016htt, Abbott:2016nmj, Abbott:2017vtc}, and the last gravitational wave detection from neutron stars \cite{TheLIGOScientific:2017qsa} have open the floodgates for a large range of possibilities in gravitational physics in the same way that the discovery of infrared radiation by William Hershel in the "`1800's"' opened the possibility of the investigation and application of the electromagnetic spectrum beyond to the visible.
It can yet be reinforced by others results from others
gravitational antennas such as  LISA,
VIRGO, and others. In a timely way, the quasinormal fundamental mode
constitutes the predominating contribution 
to such signals in  the long period of damping proper oscillations \cite{Konoplya:2011qq}. Moreover, the quasinormal modes' ringdown
phase of the signal is useful to distinguish between
black holes and other possible sources \cite{Chirenti:2017mwe} .

In this work, we shall address the gravitational wave production by LQBHs. 
Quasinormal frequencies for such black holes have been addressed in the scalar perturbations case in \cite{Chen:2011zzi, Santos:2015gja}. However, in a more realistic scenario, gravitational perturbations must be included.
Investigation on the quasinormal mode spectrum in the context of LQBHs may reveal some advantages 
front others scenarios under the experimental
point of view since the quantum corrections present in this scenario depend on 
the  dimensionless Barbero-Immirzi parameter \cite{Rovelli:2004tv}, which as it has been pointed in \cite{Sahu:2015dea} does not suffers with the problem
of mass suppression, as occurs with the parameters of other quantum gravity theories like superstring theory
or noncommutative theory. In these theories, the quantum corrections appears as proportional to $(l_{\rm qg}/M)^m$, where
$l_{\rm qg}$ is a quantum gravity motivated dimensionful parameter of the theory, $M$ is the black hole mass, and $m$ is some positive number. In this way, for large black holes, the corrections coming from quantum gravity effects are mass suppressed.

The article is organized as follows: in section \eq{sdbh}, we revise the LQBH's theory;
in section \eq{rw-formalism}, we calculate the Regge-Wheeler equation for LQBHs; in section 
\eq{qn-modes},
we calculate the quasinormal modes of the LQBHs. Section \eq{conclusions} is devoted to conclusions and perspectives. 
Throughout this paper, we have used $\hbar = c = G = 1$.

\section{Loop quantum black holes} \label{sdbh}

Some efforts in order to find out black hole solutions in LQG have been done by several authors
\cite{Kuchar:1994zk, Thiemann:1992jj, Campiglia:2007pr, Modesto:2004xx, Bengtsson:1988hm, Bojowald:1999eh, Bojowald:2004ag, Bojowald:2004si, Bojowald:2005cb, Modesto:2008im, Gambini:2013ooa}.
In this section, we shall analyze a particular solution, called self-dual, that appeared at the primary time from a simplified model of LQG consisting in symmetry reduced models corresponding to homogeneous spacetimes (see \cite{Modesto:2008im} and the references therein). 

In this way, the LQBH framework which we shall work here is delineated by a quantum gravitationally corrected Schwarzschild metric, written as

\begin{equation}
ds^{2} = - G(r)dt^{2} + F^{-1}(r)dr^{2} + H(r)d\Omega^{2} \label{self-dual-metric}
\end{equation}

\noindent with

\begin{equation}
 d\Omega^{2} = d\theta^{2} + \sin^{2}\theta d\phi^{2}\; ,
\end{equation}

\noindent where, in the equation \eq{self-dual-metric}, the metric functions are given by

\begin{eqnarray}
G(r) &=& \frac{(r-r_{+})(r-r_{-})(r+r_{*})^2}{r^{4}+a_{0}^{2}} \; , \nonumber \\
F(r) &=& \frac{(r-r_{+})(r-r_{-})r^{4}}{(r+r_{*})^{2}(r^{4}+a_{0}^{2})} \; , 
\end{eqnarray}

\noindent and

\begin{equation}
 H(r) = r^{2} + \frac{a_{0}^{2}}{r^{2}} \; , \label{h-form}
\end{equation}

\noindent where

\begin{eqnarray*}
r_{+} = 2m \;\; ; \;\; r_{-} = 2mP^{2} \; .
\end{eqnarray*}

\noindent In this situation, we have got the presence of two horizons - an event horizon localized at $r_{+}$
and a Cauchy horizon localized at $r_{-}$. 

Furthermore, we have that

\begin{eqnarray}
r_{*} = &\sqrt{r_{+}r_{-}} = 2mP\;.
\end{eqnarray}
\noindent In the definition above, $P$ is the polymeric function given by 
\begin{equation}
P = \frac{\sqrt{1+\epsilon^{2}} - 1}{\sqrt{1+\epsilon^{2}} +1} \;\; ; 
\end{equation}
where $\epsilon=\gamma\delta_b$, being $\gamma$ the Barbero-Immirzi parameter and $\delta_b$ is the polymeric parameter used in the LQG quantization techniques in order to determinate the length of the path along with the connection used to define the holonomies is integrated \cite{Modesto:2008im}. Moreover
\begin{equation}
a_{0} = \frac{A_{min}}{8\pi}\;,
\end{equation}

\noindent where $A_{min}$ is the minimal value of area in LQG. 

In the metric \eq{self-dual-metric}, r is only asymptotically the same old radial coordinate
since $g_{\theta\theta}$ is not simply $r^{2}$.
A more physical radial coordinate is obtained from the shape of the function $H(r)$ within the metric \eq{h-form}

\begin{equation}
R = \sqrt{r^{2}+\frac{a_{0}^{2}}{r^{2}}}\;\; , \label{phys-rad}
\end{equation}

\noindent in the sense that it measures the right circumferential distance.

Moreover,
the parameter $m$ within the solution is
related to the ADM mass $M$ by 

\begin{equation}
M = m(1 + P )^{2} \;. \label{mass-rel}
\end{equation}

The equation \eq{phys-rad} reveals vital aspects of the LQBH's internal structure. 
From this expression, we have got that, as $r$ decreases from $\infty$
to $0$, $R$ initially decreases from $\infty$ to $\sqrt{2 a_{0}}$ at $r= \sqrt{a_{0}}$ so will increase once 
more to $\infty$. The value of $R$ associated
with the event horizon is given by

\begin{equation}
R_{EH} = \sqrt{H(r_{+})} = \sqrt{(2m)^{2} + \Big(\frac{a_{0}}{2m}\Big)^{2} }\; . \label{r-horizon}
\end{equation}

An interesting feature of LQBHs scenario is that related to the self-duality of the 
metric \eq{self-dual-metric}. In this case, self-duality means that
if one introduces the new coordinates $\tilde{r} = a_{0}/r$ and $\tilde{t} = t r_{*}^{2}/a_{0}$, 
the metric
preserves its type. The dual radius is given by $r_{dual} = \tilde{r} = \sqrt{a_{0}}$ and corresponds 
to the smallest possible surface element.
Moreover, since the equation \eq{phys-rad} may be written as $R = \sqrt{r^{2}+\tilde{r}^{2}}$,
it is clear that the solution
contains another asymptotically flat Schwazschild region instead of
a singularity within the limit $r\rightarrow 0$. This new region corresponds to a Planck-sized wormhole.
Figure \eq{carter-penrose} shows the Carter-Penrose diagram for the LQBH.

\begin{figure}[htb]
 \centering 
 \fbox{\includegraphics[width=6cm,height=8cm]{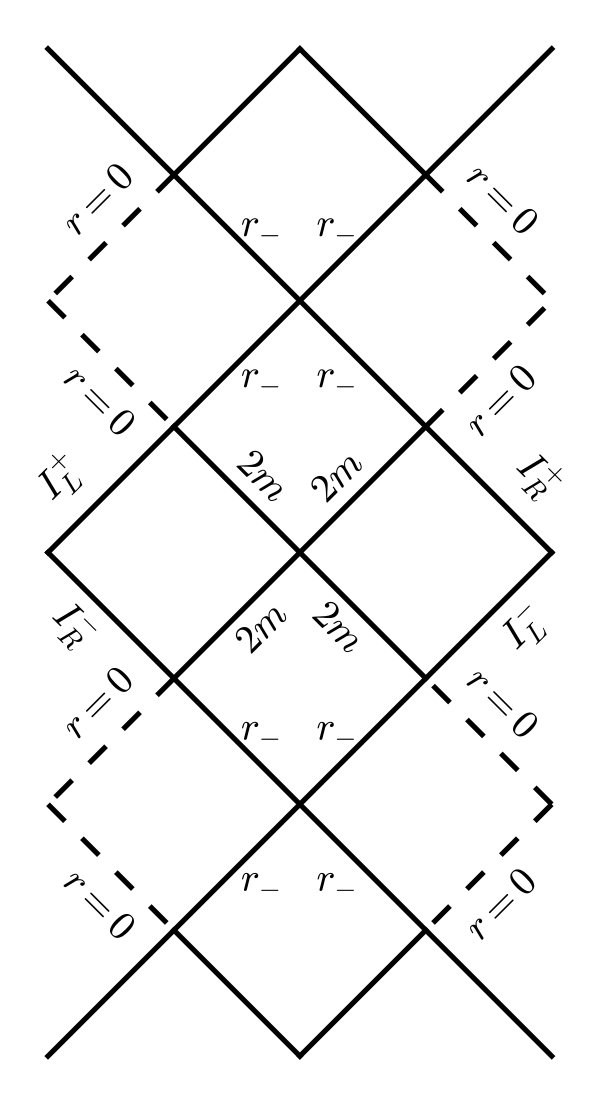}}     
\caption[Figure 2:]{Carter - Penrose diagram for the LQBH metric. The diagram has two asymptotic regions, one at infinity 
and the other near the origin, which no observer can reach in a finite time. }
 \label{carter-penrose} 
\end{figure}


In addition, the thermodynamical properties of LQBHs have been addressed 
in the references \cite{Modesto:2009ve, s.hossenfelder-prd81, Alesci:2011wn, Carr:2011pr,  s.hossenfelder-2012ca,
Silva:2012mt}.
In particular, within the reference \cite{Silva:2012mt},
such issue has been addressed by the employment of a tunneling technique
which has made possible the introduction of back-reaction effects. 
By employing the same tunneling method, in the reference \cite{Anacleto:2015mma} it has been investigated 
the influence of a Generalized Uncertainty Principle on the thermodynamics of
LQBHs. Interestingly, from the form of the LQBHs entropy, it has been found out that these objects 
could be the main constituents of a holographic version of Loop Quantum Cosmology (LQC) \cite{Silva:2015qna}. 


\section{Regge-Wheeler formalism for LQBHs} \label{rw-formalism}

In this section, we shall use a method due to 
Regge and Wheeler to investigate black hole's axial gravitational perturbations  \cite{Regge:1957td}.
In this way, we have that, if small perturbations are introduced, the resulting spacetime metric can be written as  
$g_{\mu\nu} = \tilde{g}_{\mu\nu} + h_{\mu\nu}$, where $ \tilde{g}_{\mu\nu}$ 
is the background metric and $h_{\mu\nu}$ is the spacetime perturbation. Moreover, the perturbations are much smaller than
the background. By placing our attention on the perturbation $h_{\mu\nu}$, we have that, due to spherical symmetry, 
it can be written in two parts, where one
depends on the angular coordinates through the spherical harmonics and the other one depends on
the coordinates $r$ e $t$. In addition, $h_{\mu\nu}$ can be written as a sum of a polar and an axial component. 

In this context, we can find out, in particular, that the axial component of the gravitational perturbation of the metric
\eq{self-dual-metric}, can be written, under the Regge-Wheeler gauge, as 
\cite{Rezzolla:2003ua}:

\begin{equation}
 h^{axial}_{\mu \nu} = \left[
\begin{array}{cccc}
0   & 0 & 0 & h_{0}(t,r) \\
0  & 0 & 0 & h_{1}(t,r) \\
0 & 0 & 0 & 0 \\
h_{0}(t,r) & h_{1}(t,r) & 0 & 0 \\
\end{array}
\right]\sin \theta \partial_{\theta}P_l(\cos \theta)e^{im\phi} \; .
\end{equation}

In the case of gravitational axial perturbations,  we have that  

\begin{eqnarray}\label{EEOM}
 \delta R_{\mu\nu}=  0  \;\; ,
\end{eqnarray} 

\noindent where 
\begin{eqnarray}\label{r-variation}
 \delta R_{\mu\nu}=\delta\Gamma_{\mu\alpha;\nu}^{\alpha}-\delta\Gamma_{\mu\nu;\alpha}^{\alpha}
\end{eqnarray}

From the equation \eq{EEOM}, we obtain

\begin{eqnarray}
 \delta R_{23}=\frac{1}{2}\left[-G^{-1}\frac{\partial h_0}{\partial t}+F\frac{\partial h_1}{\partial r}+
 \frac{1}{2}F'h_1+\frac{1}{2}\frac{G'}{G}Fh_1\right] \times \left\{\cos \theta \frac{\partial P_l(\cos \theta )}{\partial \theta}
 \right. \nonumber \\-
 \left. \sin \theta \frac{\partial^2P_l(\cos \theta )}{\partial \theta^2}\right\}=0,
\end{eqnarray}

\begin{eqnarray}
 \delta R_{13}=\frac{1}{2}\left[G^{-1}\left(\frac{\partial^2h_0}{\partial t \partial r}-\frac{H'}{H}
 \frac{\partial h_0}{\partial t}\right)+\left(-\frac{l(l+1)}{H}h_1-G^{-1}\frac{\partial^2h_1}{\partial t^2}+
 \frac{1}{2}F'\frac{H'}{H}h_1\right.\right. \nonumber\\
 \left.\left.+F\frac{H''}{H}h_1+\frac{1}{2}\frac{G'}{G}F\frac{H'}{H}h_1\right)\right] \times \left\{
 \sin \theta \frac{\partial_l(\cos \theta )}{\partial \theta}\right\}=0 \; .
\end{eqnarray}



In order to handle the equations above, we can observe that values of multipole indices 
for which $l<s$, are not related to
dynamical modes, corresponding to conserved quantities. In this way, we shall consider only the nontrivial radiative
multipoles with $l\geq s$. On the other hand, a gravitational perturbation corresponding to
$l=0$ will be related to a black hole mass change, and a perturbation corresponding to $l=1$ 
will be related to a displacement as well as to a change on the black hole angular momentum.
The most interesting multipole indices values are those that correspond to the cases where the wave can propagate
during a time interval large enough to be detected. In this way, only the $l \geq 2$ cases are relevant 
\cite{Frolov:1998wf}.
Since, for $l \geq 2$, we have $P_{l \geq 2} \neq 0$, we obtain the following radial equations

\begin{eqnarray}
 -G^{-1}\frac{\partial h_0}{\partial t}+F\frac{\partial h_1}{\partial r}+\frac{1}{2}F'h_1+\frac{1}{2}
 \frac{G'}{G}Fh_1=0, \label{df-1}
\end{eqnarray}
\begin{eqnarray}
 \left.-\frac{\partial^2 h_0}{\partial t \partial r}+\frac{\partial^2h_1}{\partial t^2}+\frac{H'}{H}\frac{\partial h_0}
 {\partial t}+\left[\frac{G}{H}l(l+1)-\frac{1}{2}GF'\frac{H'}{H}-\frac{GFH''}{H}-\frac{1}{2}\frac{G'FH'}{H}
 \right]h_1=0\right. \; .\label{df-2}
\end{eqnarray}

\noindent Moreover, from the equation \eq{df-1}, we have that

\begin{eqnarray}
 \frac{\partial h_0}{\partial t}&=&(GF)^{1/2}\left \{(GF)^{1/2}\frac{\partial h_1}{\partial r}+
 \frac{1}{2}(GF)^{-1/2}GF'h_1+\frac{1}{2}(GF)^{-1/2}G'Fh_1\right \} \nonumber\\
 &=&(GF)^{1/2}\frac{\partial Q(t,r)}{\partial r}=\frac{\partial Q(t,r)}{\partial x}\; ,
\end{eqnarray}

\noindent where the function $Q(t,r)$ is defined as

\begin{equation}
 Q(t,r) \equiv (GF)^{1/2}h_1(t,r), \label{q-eq}
\end{equation}

\noindent and the tortoise coordinate $x$ is given by the relation:

\begin{eqnarray}
 \frac{\partial r}{\partial x}=(GF)^{1/2}\; . \label{tortoise-eq}
\end{eqnarray}

\noindent By integrating the equation above, we obtain

\begin{eqnarray}
 x&=& r-\frac{a^{2}_{0}}{rr_{-}r_{+}}+\frac{a^{2}_{0}(r_{-}+r_{+})}{r^{2}_{-}r^{2}_{+}}\log(r)+
 \frac{(a^{2}_{0}+r^{4}_{-})}{r^{2}_{-}(r_{-}-r_{+})}
 \log(r-r_{-}) \nonumber\\
 &+&\frac{(a^{2}_{0}+r^{4}_{+})}{r^{2}_{+}(r_{+}-r_{-})}\log(r-r_{+}).
\end{eqnarray}

Now, if we substitute the equations \eq{q-eq} and \eq{tortoise-eq} in \eq{df-2} , we obtain:

\begin{eqnarray}
 -\frac{d}{dr}\frac{dQ}{dx}+(GF)^{-1/2}\frac{d^{2}Q}{dt^{2}}+\frac{H'}{H}\frac{dQ}
{dx}
 +\left[\frac{G}{H}l(l+1)-\frac{1}{2}GF'\frac{H'}{H}-GF\frac{H''}{H}\right. \nonumber \\ \left. -\frac{1}{2}G'F\frac{H'}{H}\right]
 (GF)^{-1/2}Q=0 \; ,
\end{eqnarray}

\noindent which, by using the definition

\begin{eqnarray}
\Psi(t,x(r)) \equiv \frac{Q(t,r)}{H^{1/2}},
\end{eqnarray}

\noindent can be written as

\begin{eqnarray}
 -\frac{d}{dr}\frac{d}{dx}(H^{1/2}\Psi)+\left(\frac{GF}{H}\right)^{-1/2}\frac{d^{2}\Psi}{dt^{2}}+\frac{H'}{H}
 \frac{d}{dx}(H^{1/2}\Psi)+\left[\frac{G}{H}l(l+1)-\frac{1}{2}GF'\frac{H'}{H} \right. \nonumber \\ 
 \left. -GF\frac{H''}{H}-\frac{1}{2}G'F\frac{H'}{H}\right]\left(\frac{GF}{H}\right)^{-1/2}\Psi=0\; .
\end{eqnarray}

The equation above can be rewritten as a Schrodinger-type equation which reads:

\begin{eqnarray}
 -\frac{d^{2}\Psi}{dx^{2}}+\frac{d^{2}\Psi}{dt^{2}}+V_{eff}(r(x))\Psi=0,
\end{eqnarray}

\noindent where the effective potential is given by

\begin{eqnarray}
V_{eff}(r(x))&=&\frac{r^{2}(r-r_{-})(r-r_{+})}{(r^{4}+a_{0}^{2})^{4}}
\Big\{[6r_{*}^{2} - 3(r_{+} + r_{-}) + l(l+1)(r+ r_{*})^{2}]r^{8}\nonumber\\
&+& 30r^{4}a_{0}^{2}[r(r_{+}+r_{-}-r)-r_{*}^{2}] + 2r^{4}a_{0}^{2}l(l+1)(r+ r_{*})^{2} \nonumber \\
&+& 3ra_{0}^{4}[2r-(r_{+}-r_{-})] + a_{0}^{4}(r+r_{*})^{2}l(l+1)\Big\}. \label{potential}
\end{eqnarray}

\noindent The LQBH potential behavior in relation to the $r$ and $x$ coordinates are, respectively, depicted in the figures \eq{graf1} and \eq{graf2}.

\begin{figure}[h]
\centering
\includegraphics[width=12cm, height=6.5cm]{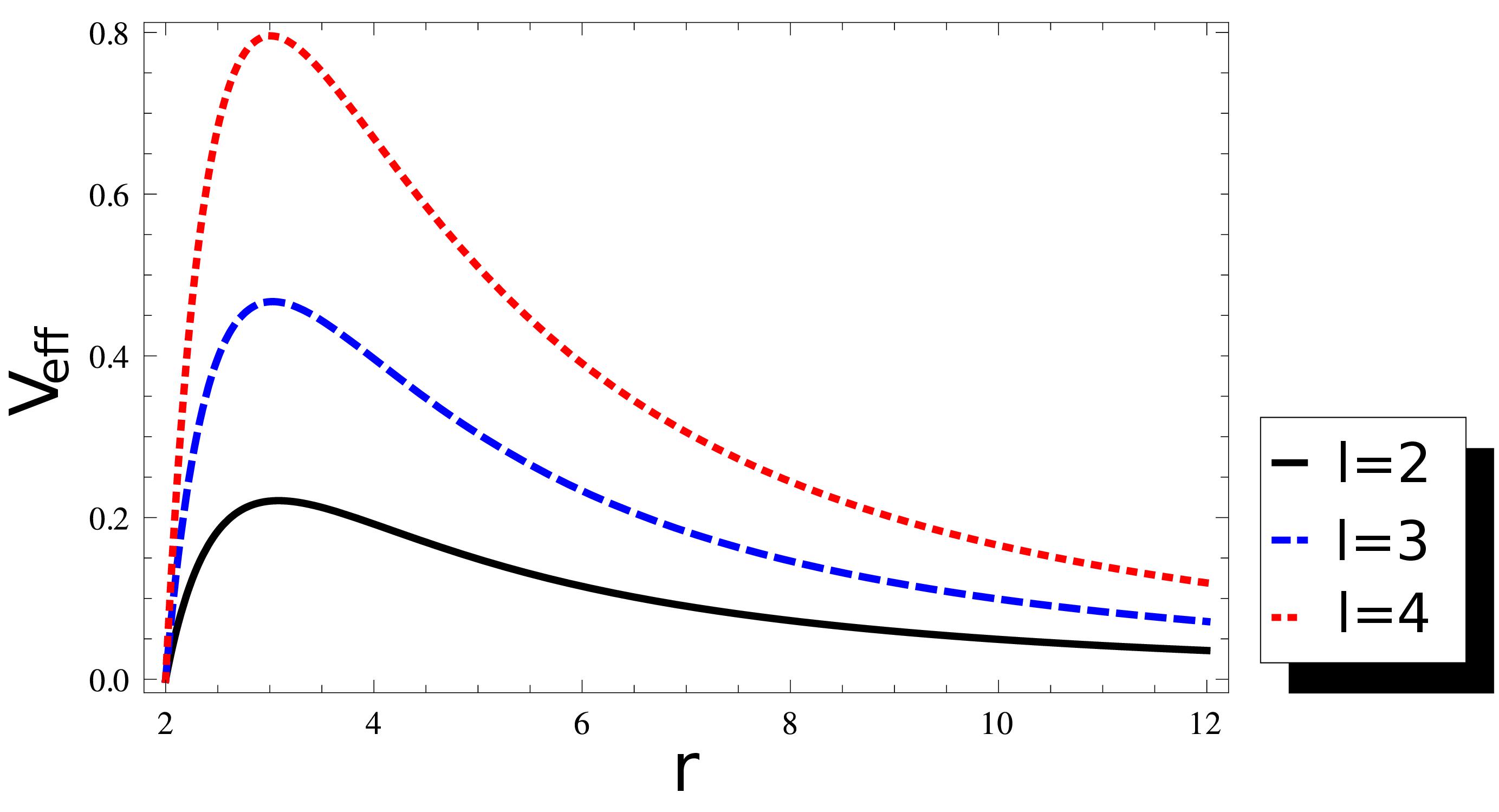}
\caption{The effective potential $V_{eff}$ for $l=2,3,4$ as a function of $r$.}
\label{graf1}
\end{figure}

\begin{figure}[h]
\centering
\includegraphics[width=12cm, height=6.5cm]{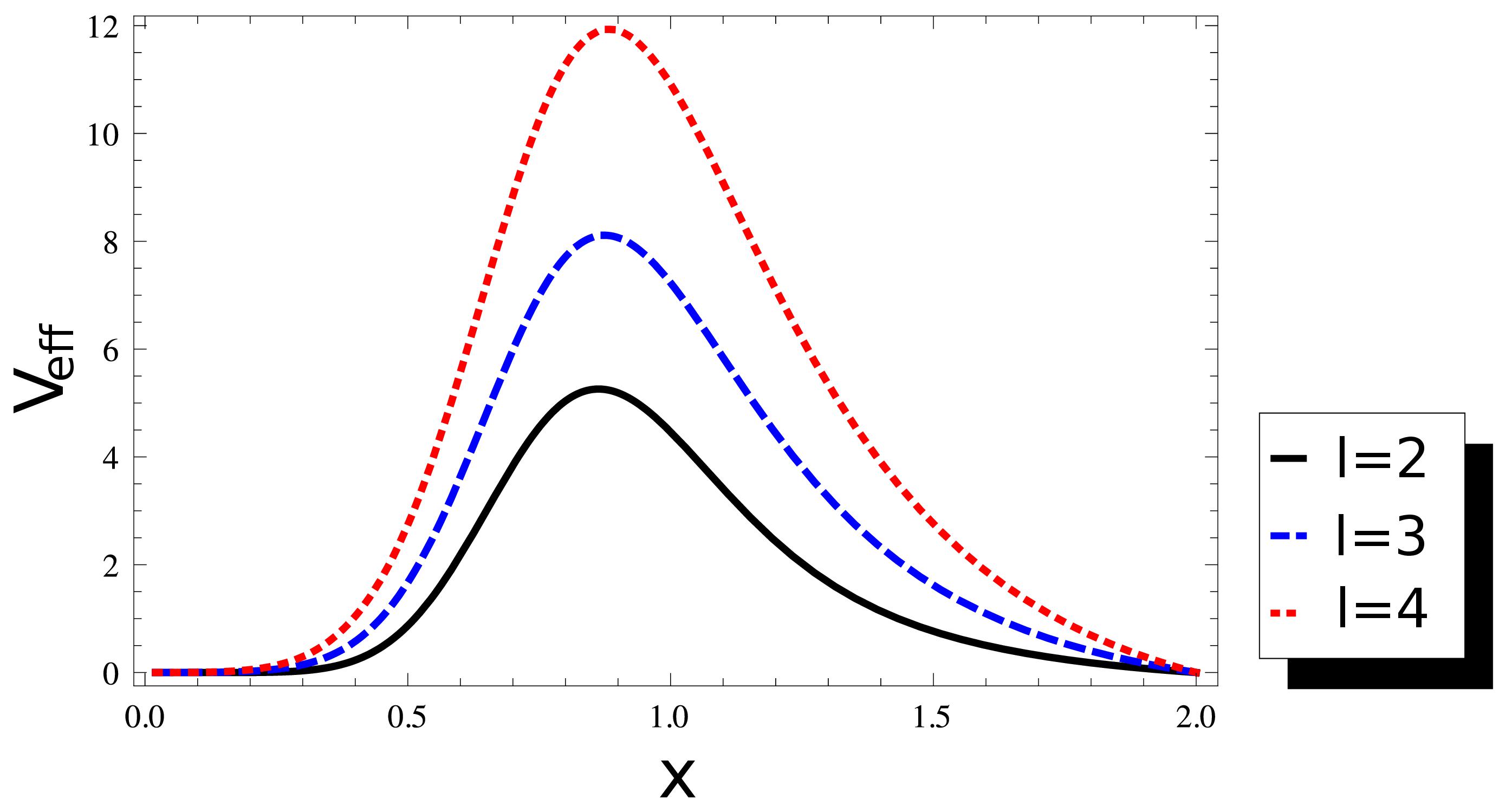}
\caption{The effective potential $V_{eff}$ for $l=2,3,4$ as a function of $x$.}
\label{graf2}
\end{figure}

As we can observe, the potential \eq{potential} contains quantum gravitational contributions to the classical potential found out by Regge and Wheeler. These contributions depend on the LQG parameters such the polymeric parameter and the minimal area value. 

In the next section, we shall analyze how the quantum gravity contributions from LQBH will affect the gravitational wave spectrum emitted by them. In order to do this, we shall analyze its quasinormal modes spectrum with the 
use of the WKB method.


\section{Quasinormal modes from LQBHs} \label{qn-modes}

In order to obtain black hole quasinormal modes, several methods have been developed, which 
date back to the beginning of $70$s.  The simplest one consists in approximating the black hole potential by 
a Poschl-Teller one whose analytical solutions are known \cite{hj.blome-pla1984}. 
Another approximation method consists in the WKB method introduced by Schutz and Will \cite{Schutz:1985zz}. Such technique is equivalent to find out the poles of the transmission coefficients of a Quantum Mechanics tunneling problem. This treatment was after improved to the third order by Iyer and Will \cite{s.iyer-prd35}, and to the sixth order by Konoplya \cite{Konoplya:2011qq}. 
Moreover, following Chandrasekhar and Detweiler \cite{Chandrasekhar:1975zza}, a shooting treatment in order to match the
asymptotic solutions at some intermediate point can be also used. Another important
method consists in the direct integration of the perturbation equation in the time domain by the use of light-cone coordinates \cite{Gundlach:1993tp}.
In addition, we have yet the continued
fraction method developed in $1985$ by Leaver \cite{Leaver:1985ax, Leaver:1990zz} and
later improved by Nollert \cite{Nollert:1993zz}. For a review of
the available techniques, we suggest \cite{Berti:2009kk} (see also \cite{Kokkotas:1999bd, Nollert:1999ji, Konoplya:2011qq} 
for other readings about QNMs).


In this section, we shall
use a WKB method due to Schutz and Will \cite{Schutz:1985zz}, and further improved by Iyer and Will \cite{s.iyer-prd35} in order to obtain the LQBH's quasinormal modes in the third order approximation.
Following this method, if one supposes that $\Psi(t,x(r))$ has a harmonic asymptotic behavior
in $t$, $\Psi \sim e^{-i\omega(t\pm x)}$, and $V_{eff}(r(x)) \rightarrow 0$ as $x \rightarrow \pm \infty$,
we obtain the following wave equation:
\begin{eqnarray}
 \frac{d^{2}\Psi}{dx^{2}}+\omega_{n}^{2}\Psi-V_{eff}(r(x))\Psi=0. \label{eq1}
\end{eqnarray}

\noindent In the equation above, the frequencies $\omega_{n}$ are determined at third order approximation by the following relation:

\begin{eqnarray}
 \omega^{2}_{n}=\left[V_{0}+\Delta \right]-i\left(n+\frac{1}{2}\right)
 \left(-2V''_{0}\right)^{1/2}\left(1+\Omega \right), \label{f-equ1}
\end{eqnarray}
\noindent where
\begin{eqnarray}
 \Delta=\frac{1}{8}\left[\frac{V_{0}^{(4)}}{V''_{0}}\right]
 \left(\frac{1}{4}+\alpha^{2}\right)-\frac{1}{288}\left(\frac{V'''_{0}}{V''_{0}}\right)^{2}\left(7+60\alpha^{2}
 \right), \label{f-equ2}
\end{eqnarray}
\begin{eqnarray}
 \Omega &=&-\frac{1}{2V''_{0}}\Big[\frac{5}{6912}\left(\frac{V'''_{0}}{V''_{0}}\right)^{4}\left(77+188\alpha^{2}
 \right)-\frac{1}{384}\left(\frac{(V'''_{0})^{2}(V^{(4)}_{0})}{(V''_{0})^{3}}\right)\left(51+100\alpha^{2}
 \right) \nonumber \\ &+&\frac{1}{2304}\left(\frac{V^{(4)}_{0}}{V''_{0}}\right)^{2}\left(65+68\alpha^{2}\right)+
 \frac{1}{288}\left(\frac{V'''_{0}V^{(5)}_{0}}{(V''_{0})^{2}}\right)\left(19+28\alpha^{2}\right) \nonumber \\
 &-&  \frac{1}{288}\left(\frac{V^{(6)}_{0}}{V''_{0}}\right)\left(5+4\alpha^{2}\right) \Big]\; . \label{f-equ3}
\end{eqnarray}

\noindent In the relations above, $\alpha=n+\frac{1}{2}$ and $V_0^{(n)}$ denotes the $n$-order derivative of the potential on the maximum $x_0$ of the 
potential.

By the use of the potential \eq{potential} in the relations \eq{eq1}, \eq{f-equ1}, \eq{f-equ2}, and \eq{f-equ3}, we can find out the quasinormal frequencies
for a LQBH, which have been depicted in the tables \eq{table1}, \eq{table2} and \eq{table3}. One may find the Schwarzchild case in Ref.~ \cite{Konoplya:2004ip}.

\begin{table}[!h]
\begin{center}
\scriptsize
\begin{tabular}{|c|c|c|c|c|}
\hline
{\bf $P$} & {\bf $\omega_{0}$} & {\bf $\omega_{1}$} & {\bf $\omega_{2}$} \\
\hline
{\bf $0.1$} & {\bf $0.3987880 - 0.0928877 i$} & {\bf $0.3652110 - 0.2869660 i$} & 
{\bf $0.3107540 - 0.4942550 i$} \\
\hline
{\bf $0.2$} & {\bf $0.4321490 - 0.0941480 i$} & {\bf $0.4057030 - 0.2896530 i$} & 
{\bf $0.3641310 - 0.4962680 i$}\\
\hline
{\bf $0.3$} & {\bf $0.4620600 - 0.0935545 i$} & {\bf $0.4403830 - 0.2866220 i$} &
{\bf $0.4064410 - 0.4892180 i$}\\
\hline
{\bf $0.4$} & {\bf $0.4872150 - 0.0907627 i$} & {\bf $0.4694450 - 0.2769010 i$} & 
{\bf $0.4411340 - 0.4709230 i$}\\
\hline
{\bf $0.5$} & {\bf $0.5061020 - 0.0855586 i$} & {\bf $0.4920720 - 0.2599850 i$} & 
{\bf $0.4690900 - 0.4405170 i$}\\
\hline
{\bf $0.6$} & {\bf $0.5169450 - 0.0777077 i$} & {\bf $0.5064670 - 0.2352620 i$} & 
{\bf $0.4886680 - 0.3971150 i$}\\
\hline
{\bf $0.7$} & {\bf $0.5177740 - 0.0669562 i$} & {\bf $0.5105030 - 0.2020610 i$} & 
{\bf $0.4975960 - 0.339798 i$}\\
\hline
{\bf $0.8$} & {\bf $0.5066550 - 0.0528541 i$} & {\bf $0.5023270 - 0.1591170 i$} & 
{\bf $0.4943430 - 0.2667300 i$}\\
\hline
\end{tabular}
\caption{The first LQBH's quasinormal modes for $l = 2$.}
\label{table1}
\end{center}
\end{table}

\begin{table}[!h]
\begin{center}
\scriptsize
\begin{tabular}{|c|c|c|c|}
\hline
{\bf $P$} & {\bf $\omega_{0}$} & {\bf $\omega_{1}$} & {\bf $\omega_{2}$} \\
\hline
{\bf $0.1$} & {\bf $0.6358020 - 0.0984560 i$} & {\bf $0.6160010 - 0.2989190 i$} & 
{\bf $0.5814460 - 0.5068190 i$}\\
\hline
{\bf $0.2$} & {\bf $0.6773290 - 0.0986630 i$} & {\bf $0.6601870 - 0.2990880 i$} & 
{\bf $0.6302840 - 0.5061190 i$}\\
\hline
{\bf $0.3$} & {\bf $0.7134190 - 0.0970203 i$} & {\bf $0.6988960 - 0.2936260 i$} &
{\bf $0.6734060 - 0.4958480 i$}\\
\hline
{\bf $0.4$} & {\bf $0.7422610 - 0.0933003 i$} & {\bf $0.7303510 - 0.2819000 i$} & 
{\bf $0.7092190 - 0.4750270 i$}\\
\hline
{\bf $0.5$} & {\bf $0.7618620 - 0.0872996 i$} & {\bf $0.7524960 - 0.2633440 i$} & 
{\bf $0.7356190 - 0.4427890 i$}\\
\hline
{\bf $0.6$} & {\bf $0.7700050 - 0.0788193 i$} & {\bf $0.7630090 - 0.2374000 i$} & 
{\bf $0.7501440 - 0.3982930 i$}\\
\hline
{\bf $0.7$} & {\bf $0.7642560 - 0.0676263 i$} & {\bf $0.7593800 - 0.2034110 i$} & 
{\bf $0.7502010 - 0.3405640 i$}\\
\hline
{\bf $0.8$} & {\bf $0.7422370 - 0.0532427 i$} & {\bf $0.7393210 - 0.1599840 i$} & 
{\bf $0.7337250 - 0.2674180 i$}\\
\hline
\end{tabular}
\caption{The first LQBH's quasinormal modes for $l = 3$.}
\label{table2}
\end{center}
\end{table}

\begin{table}[!h]
\begin{center}
\scriptsize
\begin{tabular}{|c|c|c|c|}
\hline
{\bf $P$} & {\bf $\omega_{0}$} & {\bf $\omega_{1}$} & {\bf $\omega_{2}$} \\
\hline
{\bf $0.1$} & {\bf $0.8558630 - 0.1003000 i$} & {\bf $0.8411430 - 0.3028960 i$} & 
{\bf $0.8141540 - 0.5102980 i$}\\
\hline
{\bf $0.2$} & {\bf $0.9069660 - 0.1001910 i$} & {\bf $0.8940940 - 0.3023160 i$} & 
{\bf $0.8704570 - 0.5086910 i$}\\
\hline
{\bf $0.3$} & {\bf $0.9506180 - 0.0982154 i$} & {\bf $0.9396740 - 0.2960930 i$} &
{\bf $0.9194890 - 0.4975590 i$}\\
\hline
{\bf $0.4$} & {\bf $0.9846000 - 0.0941816 i$} & {\bf $0.9756130 - 0.2836770 i$} & 
{\bf $0.9589180 - 0.4760390 i$}\\
\hline
{\bf $0.5$} & {\bf $1.0064800 - 0.0879106 i$} & {\bf $0.9993990 - 0.2645530 i$} & 
{\bf $0.9861180 - 0.4433250 i$}\\
\hline
{\bf $0.6$} & {\bf $1.0135200 - 0.0792181 i$} & {\bf $1.0082300 - 0.2381920 i$} & 
{\bf $0.9981560 - 0.3985940 i$}\\
\hline
{\bf $0.7$} & {\bf $1.0027500 - 0.0678756 i$} & {\bf $0.9990480 - 0.2039320 i$} & 
{\bf $0.9919080 - 0.3408240 i$}\\
\hline
{\bf $0.8$} & {\bf $0.9712540 - 0.0533938 i$} & {\bf $0.9690330 - 0.1603310 i$} & 
{\bf $0.9647010 - 0.2676880 i$}\\
\hline
\end{tabular}
\caption{The first LQBH's quasinormal modes for $l = 4$.}
\label{table3}
\end{center}
\end{table}

The presented results imply in a larger oscillation, as well as a slower damping as the polymeric parameter increases. 
In this way, the LQBHs decay slower as the quantum gravity contribution becomes more significant. Moreover, as we can 
see, LQBHs are stable under axial perturbations
due to the negative sign of the imaginary part of frequencies for several values of the polymeric parameter.

\begin{figure}[!htb]
\centering
\subfloat[Real]{
\includegraphics[width=12cm, height=6.5cm]{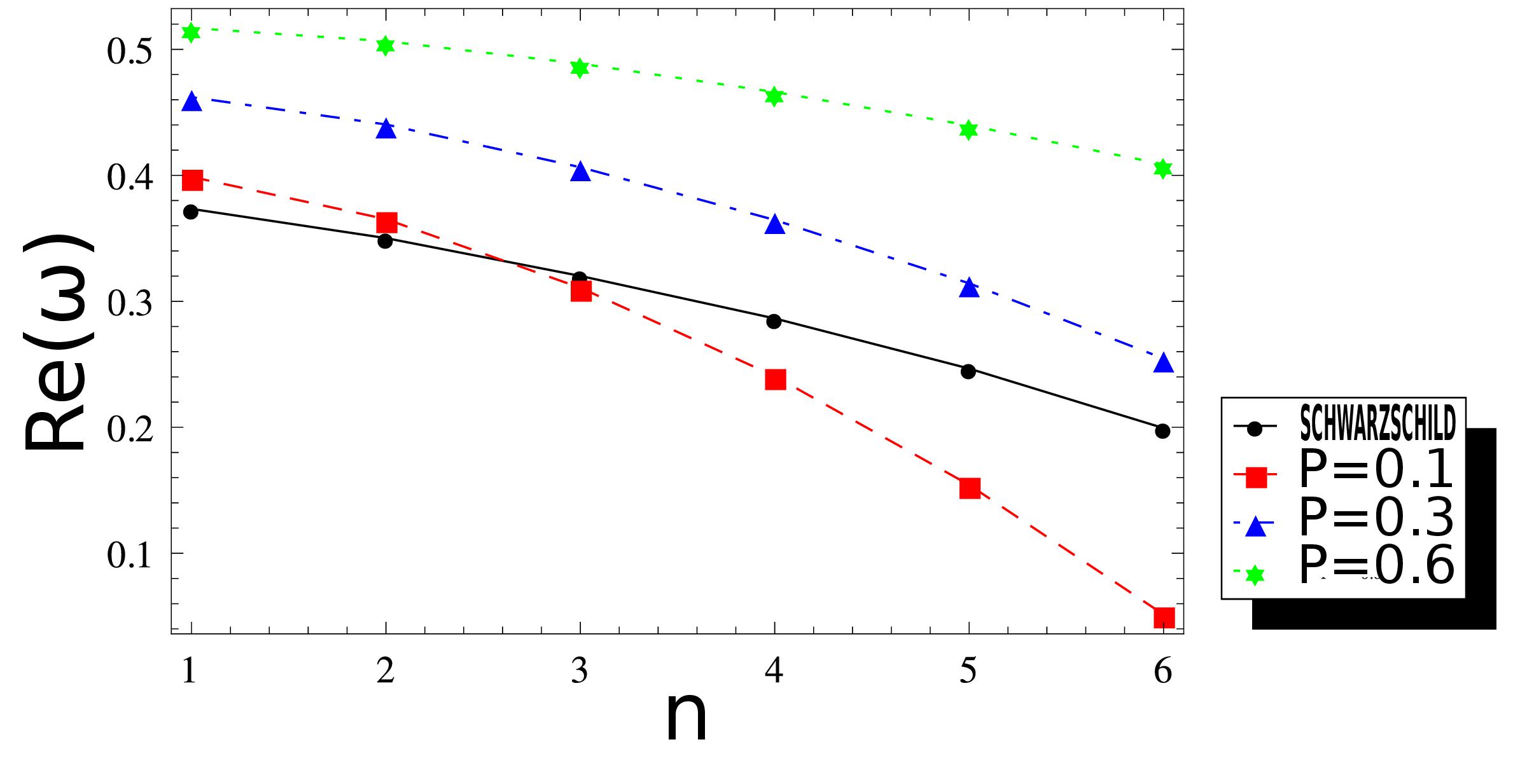}
}
\quad 
\subfloat[Imaginary]{
\includegraphics[width=12cm, height=6.5cm]{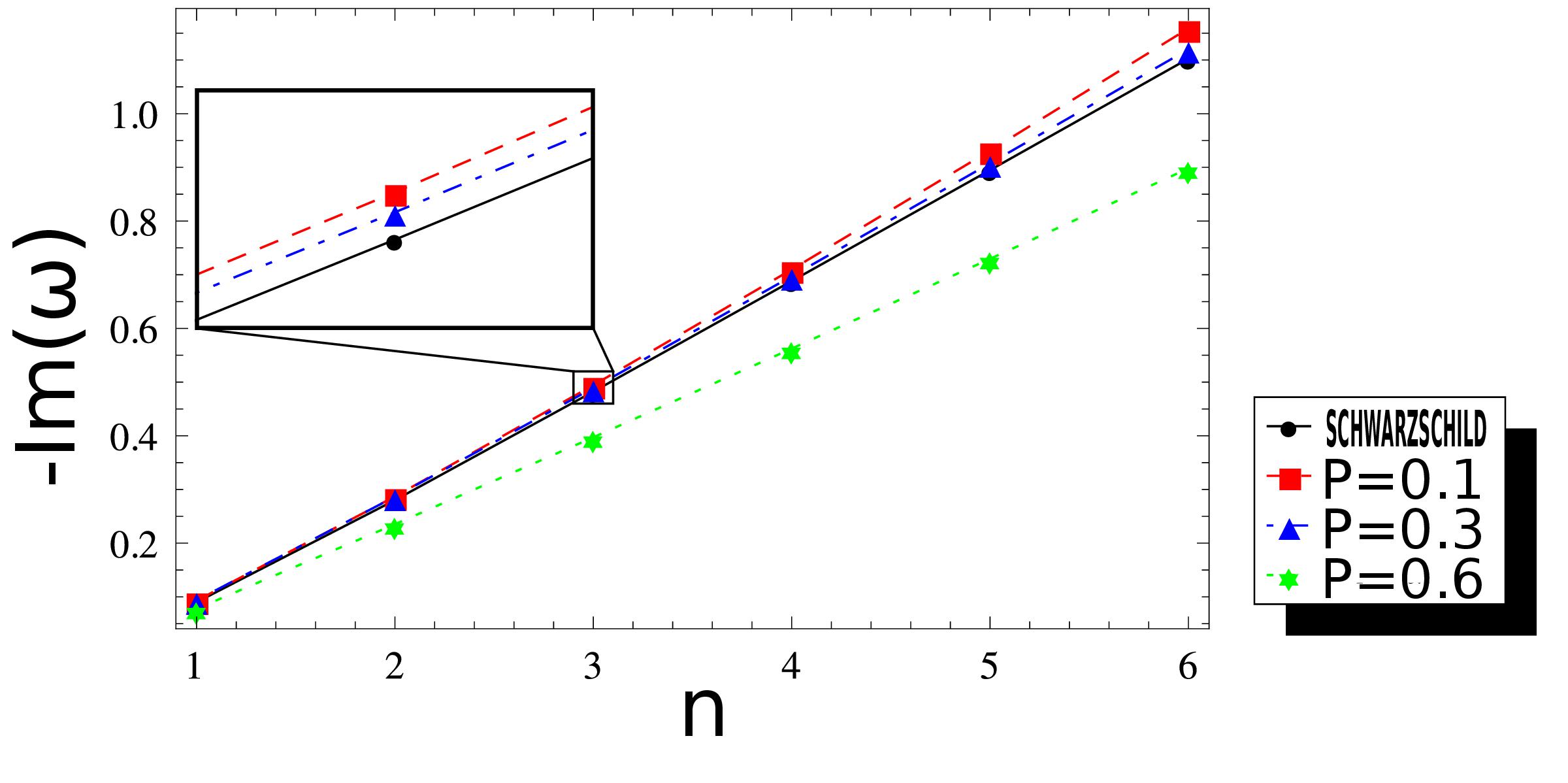}
}
\caption{Behavior of the quasinormal modes, (a) real part and (b) imaginary part of $\omega$ for $l=2$ and $P=0.1, 0.3, 0.6$.}
\label{graf3}
\end{figure}

\begin{figure}[!htb]
\centering
\subfloat[Real]{
\includegraphics[width=12cm, height=6.5cm]{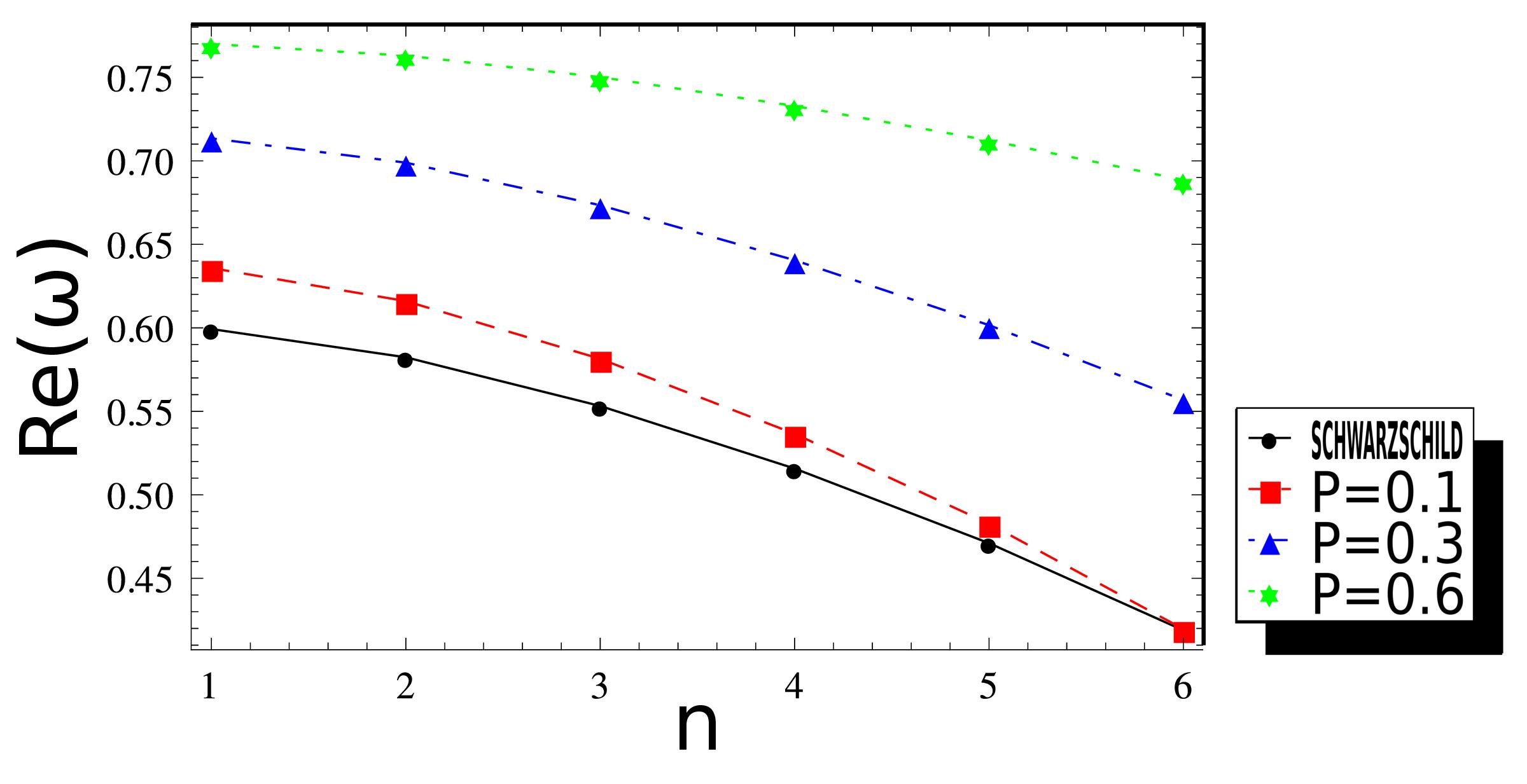}
}
\quad 
\subfloat[Imaginary]{
\includegraphics[width=12cm, height=6.5cm]{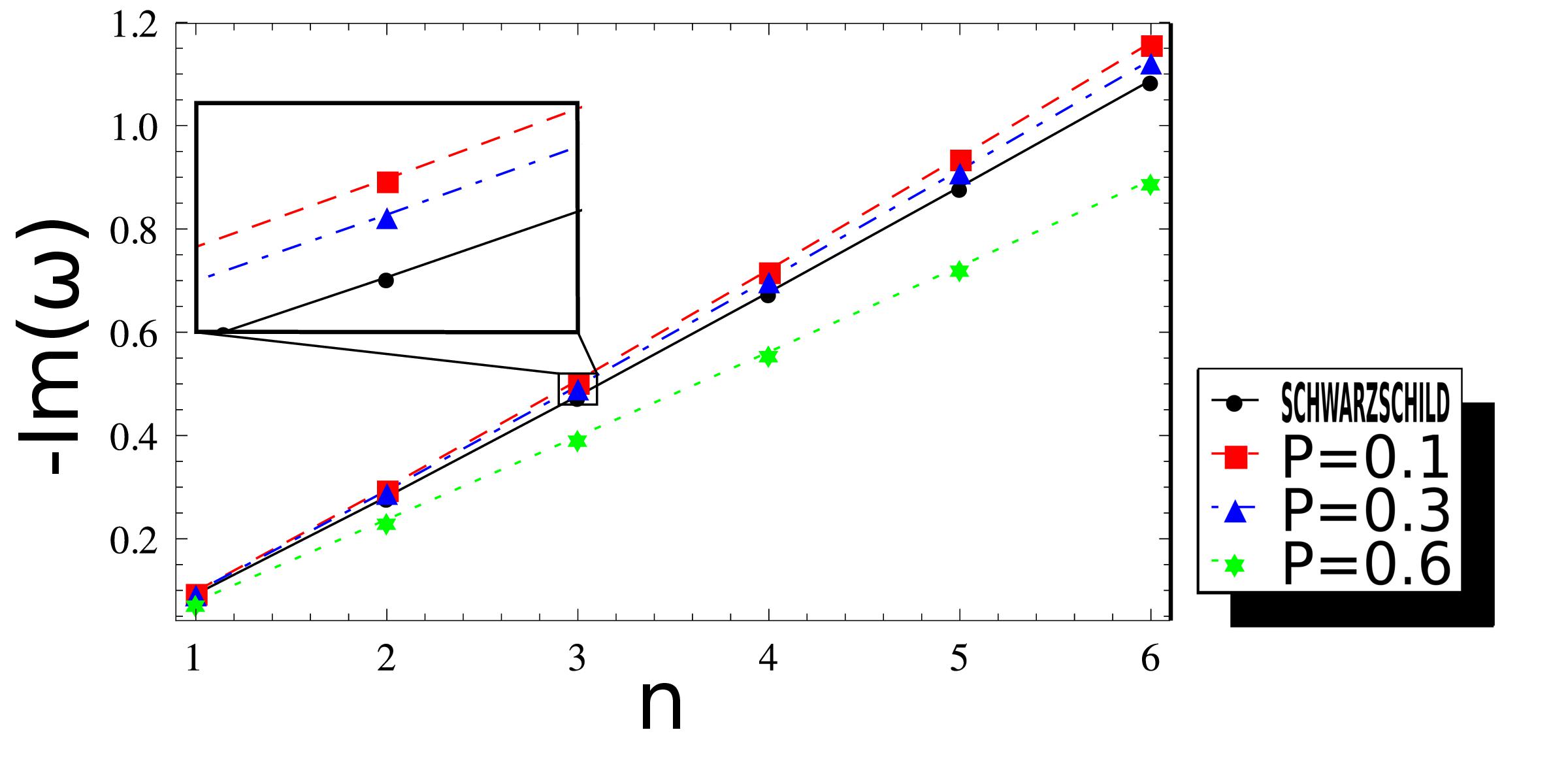}
}
\caption{Behavior of the quasinormal modes, (a) real part and (b) imaginary part of $\omega$ for $l=3$ and $P=0.1, 0.3, 0.6$.}
\label{graf4}
\end{figure}

\begin{figure}[!htb]
\centering
\subfloat[Real]{
\includegraphics[width=12cm, height=6.5cm]{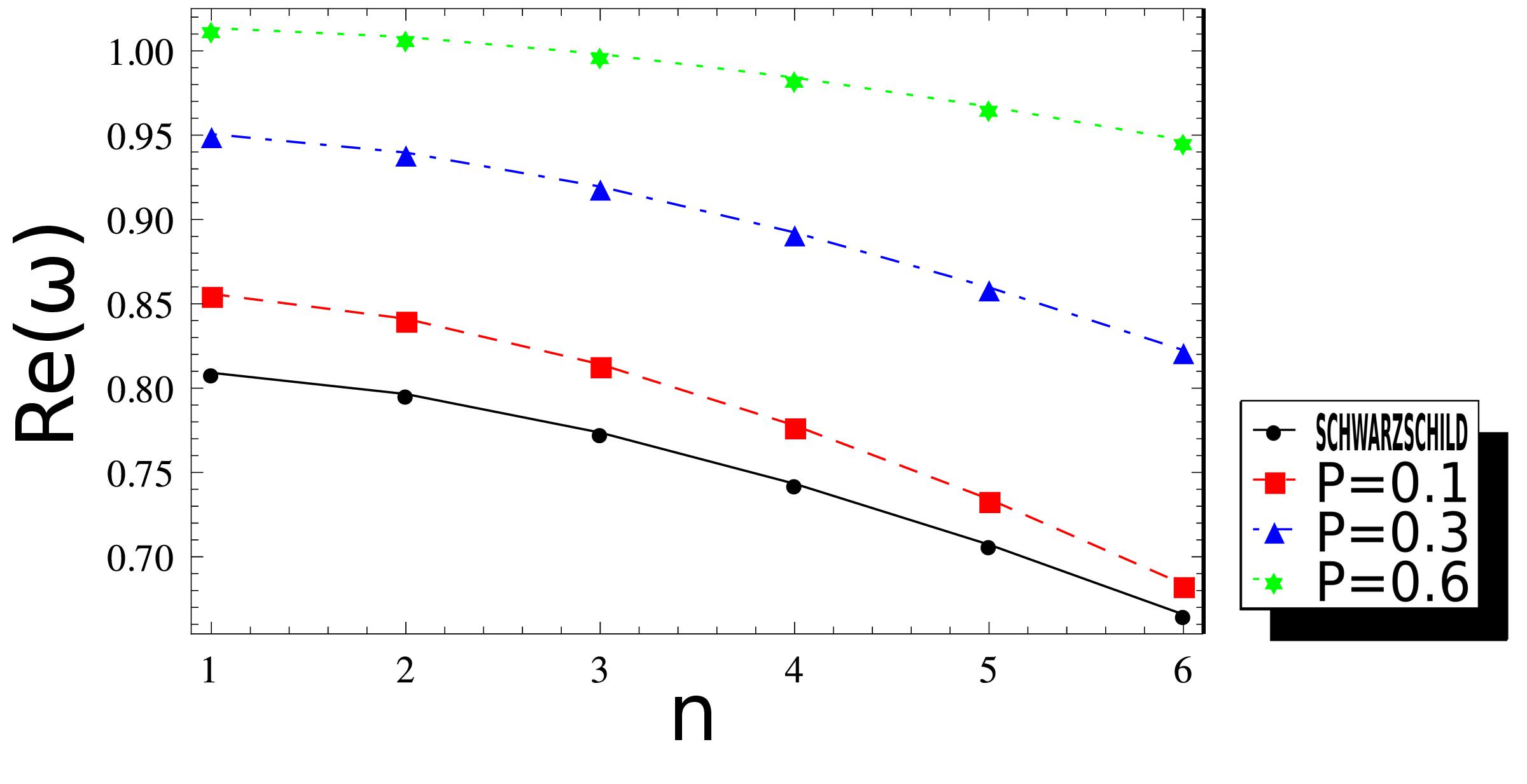}
}
\quad 
\subfloat[Imaginary]{
\includegraphics[width=12cm, height=6.5cm]{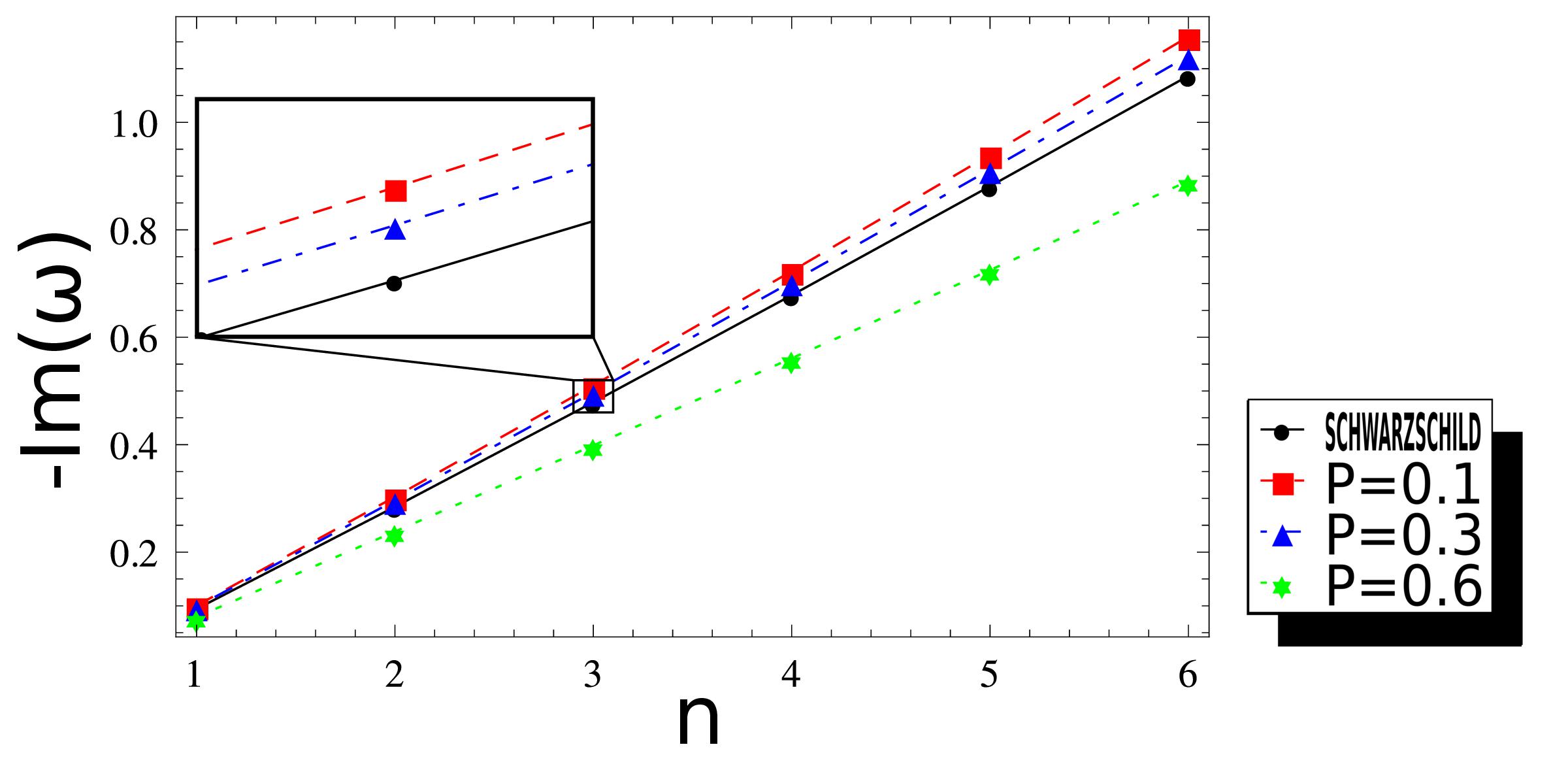}
}
\caption{Behavior of the quasinormal modes, (a) real part and (b) imaginary part of $\omega$ for $l=4$ and $P=0.1, 0.3, 0.6$.}
\label{graf5}
\end{figure}

In order to visualize, in a better way, the effects of the quantum gravity corrections present in the LQBH scenario on the black hole quasinormal spectrum, we have depicted in the graphics \eq{graf3}, \eq{graf4} and \eq{graf5}
the behavior of the LQBH quasinormal modes,  as a function of $n$,  for different values of the polymeric function $P$. In this way, we have (a) real part and (b) imaginary part of $\omega_{n}$ for $l=2,3,4$, by consideration of the following values of the polymeric function: $P=0.1$, $P=0.3$ and $P=0.6$. For the sake of comparison it was also depicted the behavior of the quasinormal spectrum of a Schwarzschild black hole.

Moreover, it is also interesting to plot the frequencies in the complex plane.
In this way, in figure \eq{graf6} this is done by considering three families of multipoles
$l = 2, 3, 4$. Looking at the right side of the figure, we conclude
that the frequency curves are moving clockwise as $P$ grows, which consists in the inverse effect we have in the case of scalar perturbations \cite{Chen:2011zzi, Santos:2015gja}. Such twisting effect
becomes more apparent for larger values of the angular quantum number. We also present the behavior for 
the Schwarzschild black hole.

\begin{figure}[!htb]
\centering
\subfloat[l = 2]{
\includegraphics[width=12cm, height=6cm]{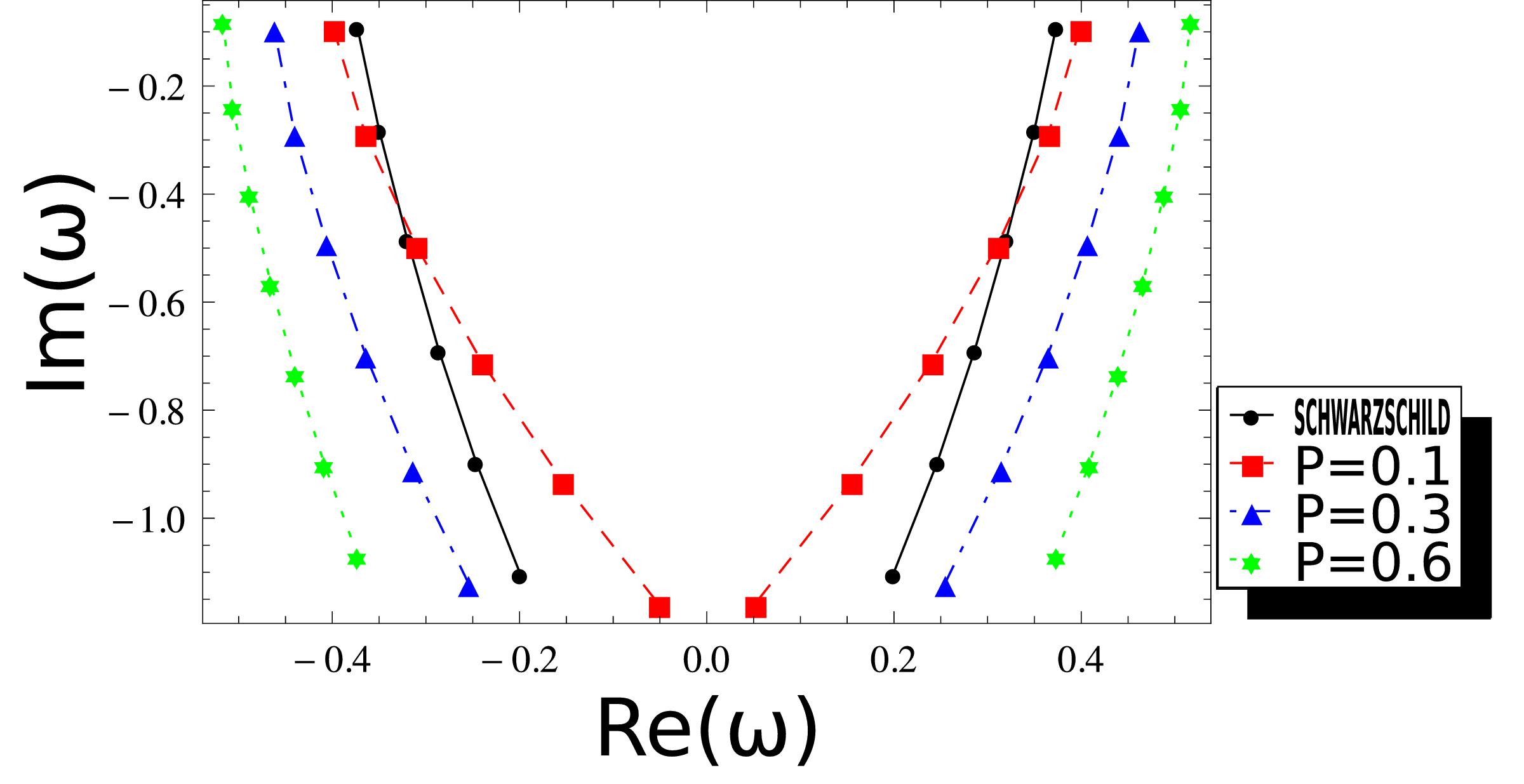}
\label{Graf_Freq_P01_Re}
}
\quad 
\subfloat[l = 3]{
\includegraphics[width=12cm, height=6cm]{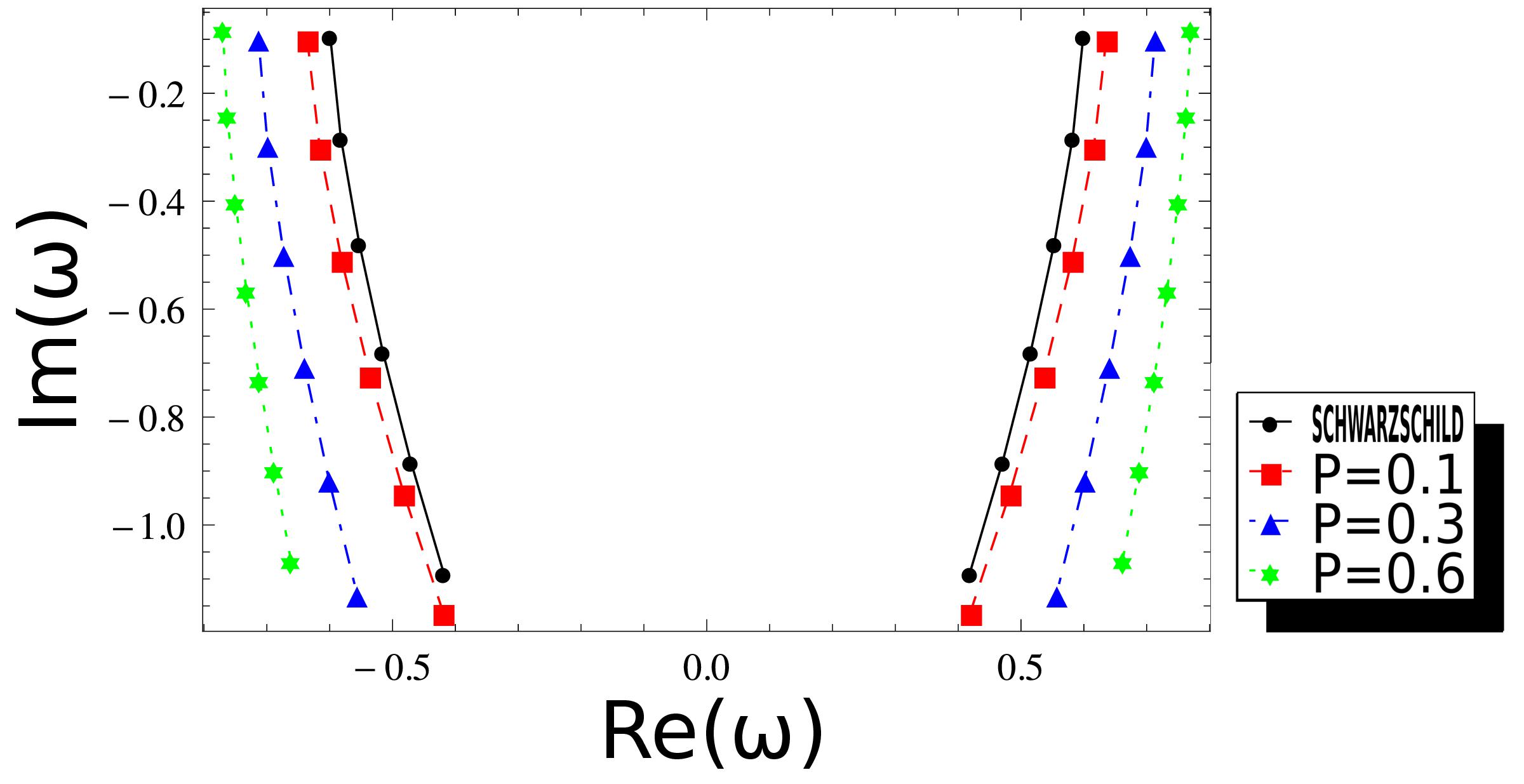}
\label{Graf_Freq_P01_Im}
}
\quad 
\subfloat[l = 4]{
\includegraphics[width=12cm, height=6cm]{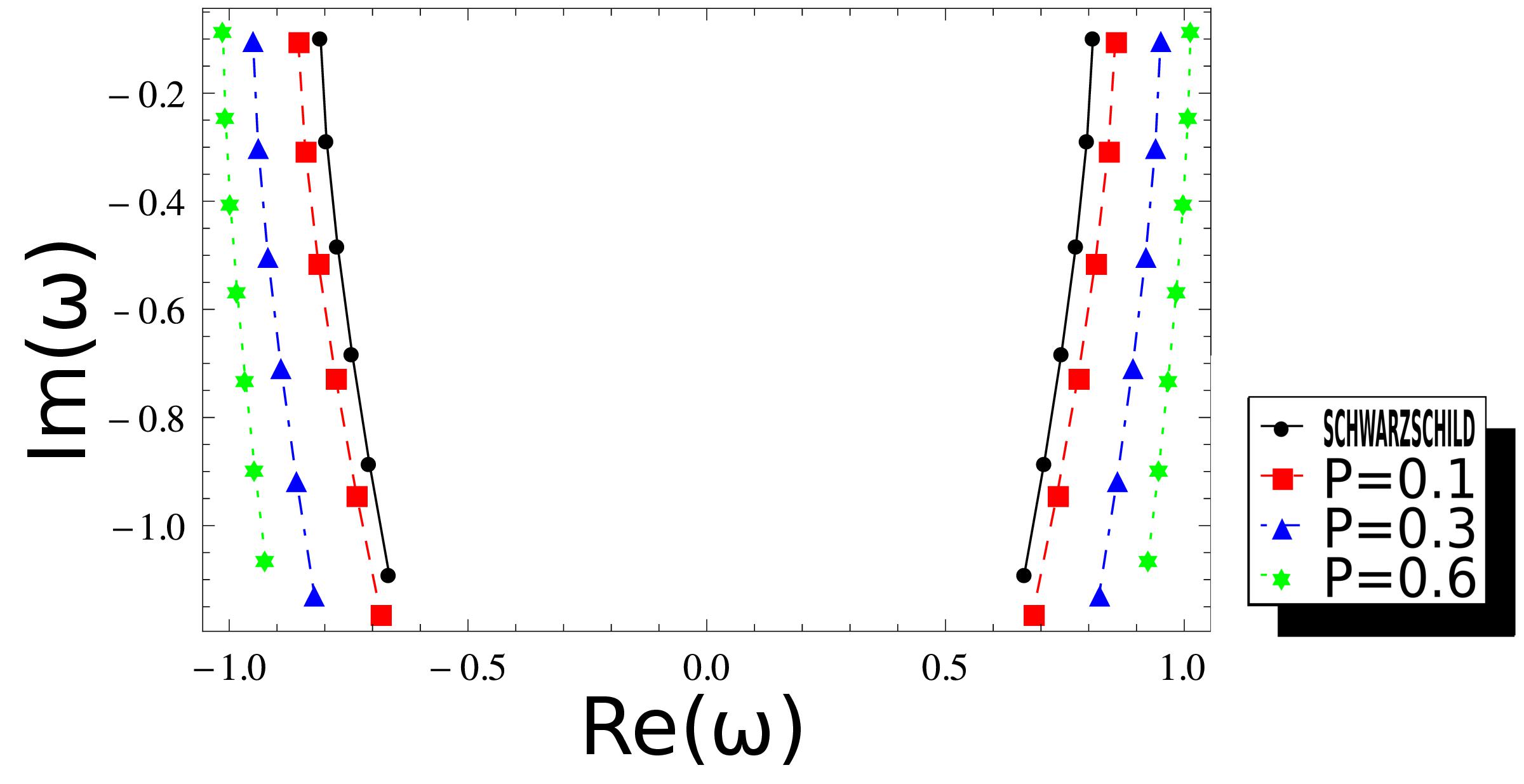}
\label{Graf_Freq_P01_Im}
}
\caption{Imaginary $\times$ Real part of $\omega$ for (a) $l=2$, (b) $l=3$, (c) $l=4$  and $P=0.1$, $P=0.3$, $P=0.6$.}
\label{graf6}
\end{figure}

From the results found out, LQBHs quasinormal modes depend directly 
from the polymeric parameter, where
a larger oscillation, as well as a slower damping occurs as the value
of the polymeric parameter increases. As a result,
the LQBHs decay slower when the quantum gravity contribution becomes more important.


However, from the recent gravitational wave observations \cite{Abbott:2016blz, TheLIGOScientific:2016htt, Abbott:2016nmj, Abbott:2017vtc, TheLIGOScientific:2017qsa}, we have that, for macroscopic black holes, with mass of the order of  the gravitational wave spectrum must agree with that given by classical General Relativity theory. In this way, for such black holes, one should not expect that quantum corrections would result in significant changes to the quasinormal radiation spectum.

In order to face up such observational facts with our results, we have multiplied the quasinormal modes by the factor $2\pi(5.142 KHz) M_{\odot}/M_{BH}$, in order to include the information about the black hole mass \cite{Konoplya:2011qq}. The results have been plotted in the figure \eq{graf7}, for the case where $l=2$, and $n= 0$, for black hole masses in the same range of that investigated in the recent observations \cite{Abbott:2016blz, TheLIGOScientific:2016htt, Abbott:2016nmj, Abbott:2017vtc, TheLIGOScientific:2017qsa}. From such results, since the polymeric function does not suffers with the mass suppression problem, we have that the recent gravitational wave observations impose a limitation to our model, where it remains valid only for $P \leq 0.1$. Such result could be used in order to impose a limitation in the value of the Barbero-Immirzi parameter, in the context of LQG.

\begin{figure}[!htb]
\centering
\subfloat[Real part]{
\includegraphics[width=12cm, height=6cm]{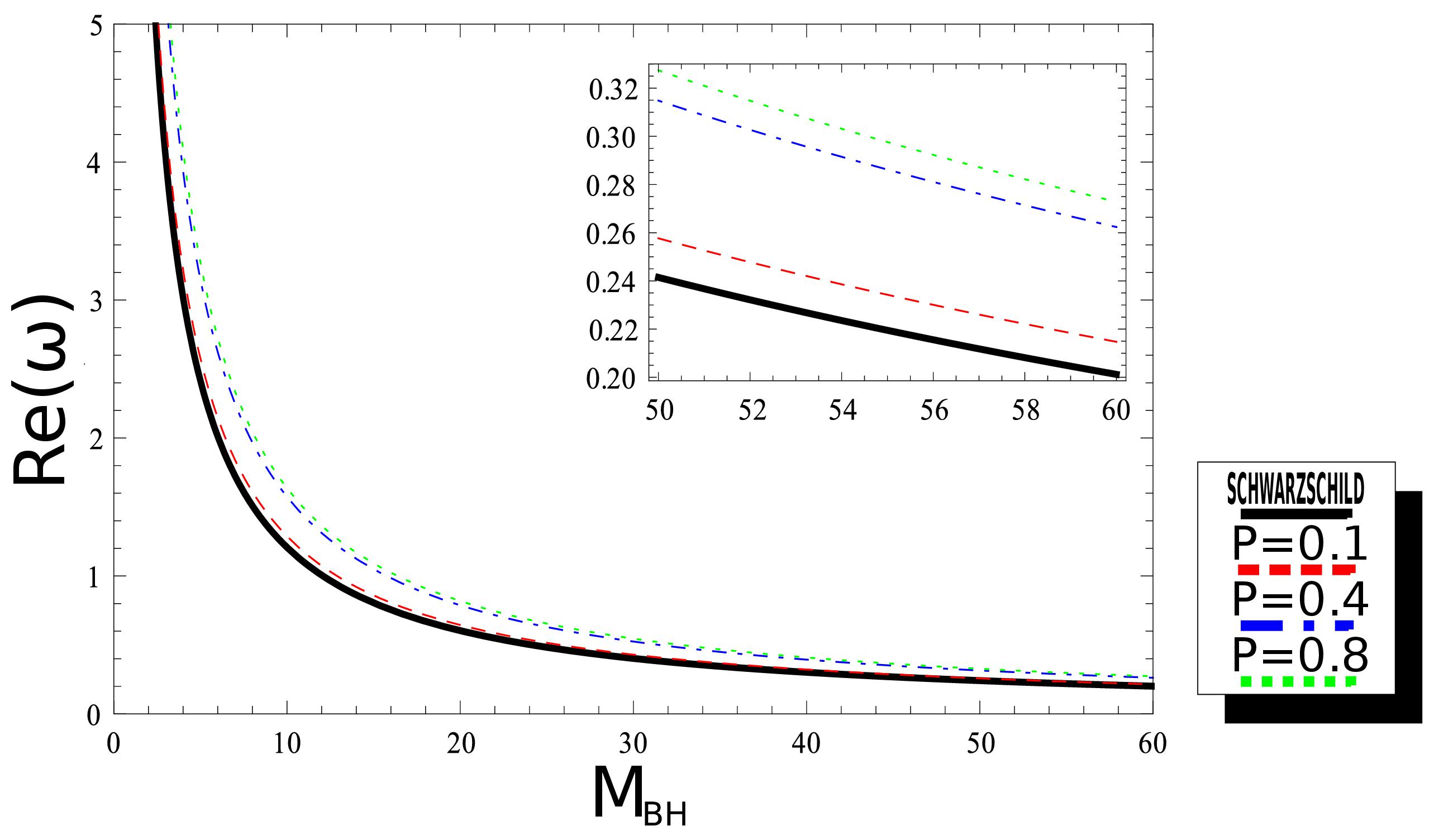}
}
\quad 
\subfloat[Imaginary part]{
\includegraphics[width=12cm, height=6cm]{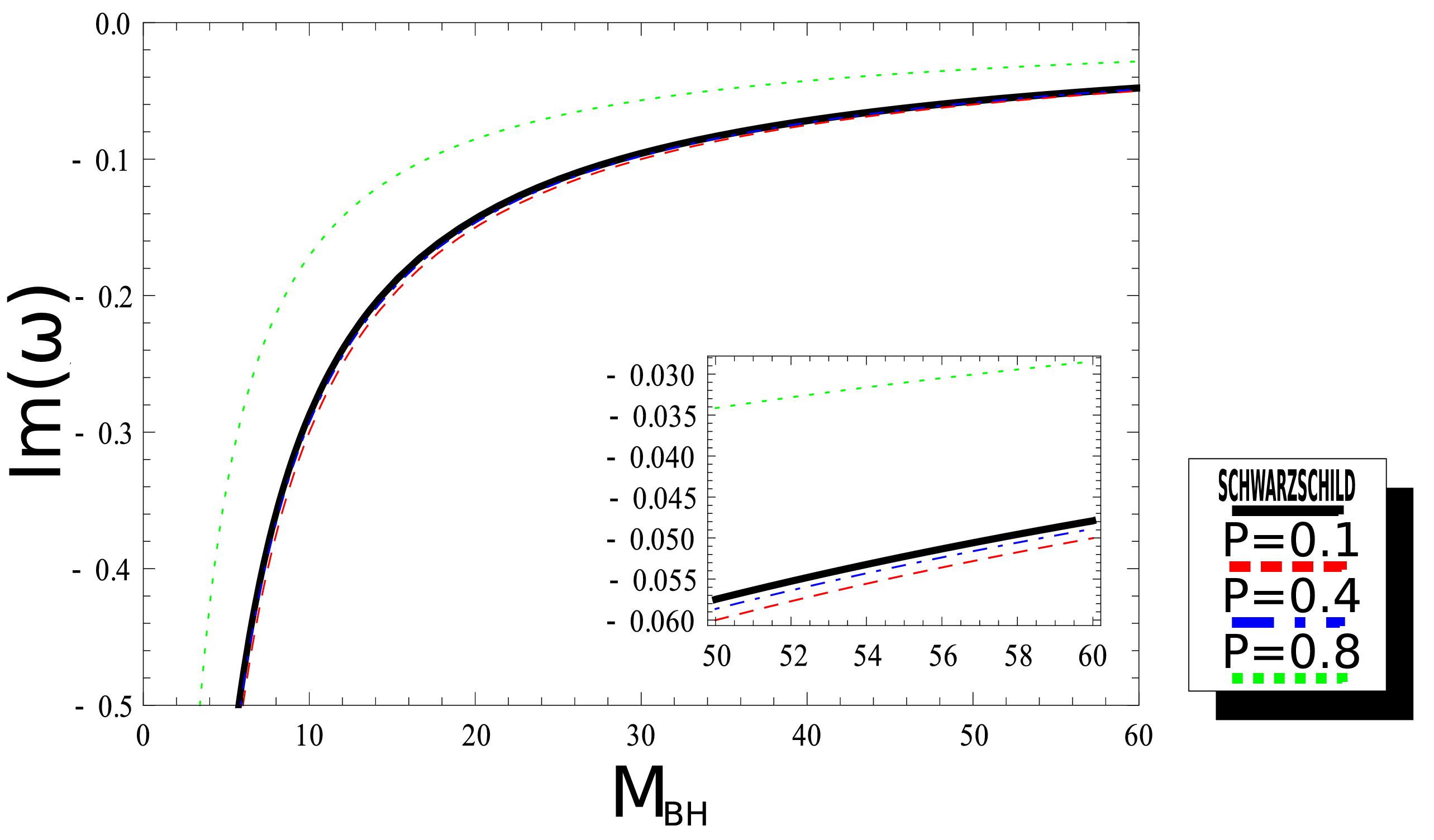}
}
\caption{Imaginary and Real part of $\omega$ for $l=2$, $n = 0$, and $P=0.1$, $P=0.4$, $P=0.8$.}
\label{graf7}
\end{figure}

\section{Conclusions and perspectives} \label{conclusions}

Black holes offer a scenario where quantum fluctuations of spacetime may appear and can be a good laboratory in order to verify the validity of predictions coming from different approaches that have arisen as candidates for a theory of quantum gravity like superstring theory and LQG.
In particular, LQBHs provide us with a way to investigate quantum gravity corrections from LQG. 

Due to its thermodynamical properties, the importance of LQBHs has been shown to go beyond the simple verification of LQG predictions to the gravitational field produced by a black hole, but it extends to the discussion of other relevant themes like the problem of dark matter and the problem of the initial state of the cosmos. It is because, LQBHs in a different way from classical Schwarzschild black holes must get thermodynamical equilibrium with radiation due to the anomalous behavior of its temperature near the Planck scale \cite{Modesto:2009ve}. In this way, LQBHs have the necessary thermodynamical stability in order to be conceived as possible candidates to dark matter \cite{Modesto:2009ve, Aragao:2016vuy}. Moreover, from the form of the entropy-area relation associated with LQBHs, they have been pointed as the building blocks of a holographic description of loop quantum cosmology \cite{Silva:2015qna}. 

Gravitational waves, on the other hand have opened the doors to a new world in physics and may establish a bridge between the quantum gravity world revealed by a black hole and experimental investigations in gravity.  In this way, motivated by the new perspectives opened by the detection of gravitational waves, we have investigated the stability of LQBHs under axial gravitational perturbations and its quasinormal spectrum.

In order to investigate how LQG corrections to a black hole scenario can influence in the gravitational wave emission, we have obtained the LQBH's quasinormal frequencies by the use of the WKB approach. The Regge-Wheeler potential corresponding to the gravitational perturbations around the LQBH was obtained and the related Regge-Wheeler equation has been solved.  After,
the quasinormal frequencies have been obtained by the use of $3th$-order WKB approximation, demonstrating that such black holes are stable under axial perturbations.

For the aforementioned importance of LQBHs as candidates to dark matter and to construct a holographic version of
LQC, it is crucial that they should be stable under gravitational perturbations. Such stability has been met in the results found out in the present paper.
Additional contributions can be obtained, beyond the
$3^{\textrm{th}}$ order approximation, by the use of the methods developed by Konoplya \cite{Konoplya:2004ip} in a forthcoming analysis.

From the results found out, LQBHs quasinormal modes depend directly 
from the polymeric parameter.
However, from the recent gravitational wave observations \cite{Abbott:2016blz, TheLIGOScientific:2016htt, Abbott:2016nmj, Abbott:2017vtc, TheLIGOScientific:2017qsa}, for macroscopic black holes, the gravitational wave spectrum must agree with that given by classical General Relativity theory. In this way, for such black holes, one should not expect that quantum corrections would result in significant changes to the quasinormal radiation spectum.

By analyzing our results for different black holes masses, we have found out a limitation to our model where
it remains valid only for $P \leq 0.1$.
Others black hole models in the context of LQG can be investigated in future works. In fact some efforts have been done in order to obtain a consistent black hole solution from LQG \cite{Kuchar:1994zk, Thiemann:1992jj, Campiglia:2007pr, Modesto:2004xx, Bengtsson:1988hm, Bojowald:1999eh, Bojowald:2004ag, Bojowald:2004si, Bojowald:2005cb, Modesto:2008im, Gambini:2013ooa}. (For a review on this issue, see \cite{Perez:2017cmj}. For some more recent works on this issue, see \cite{Ashtekar:2018lag, Ashtekar:2018cay}). The analysis of the gravitational wave spectrum from such solutions can be used as an additional criteria in order to verify, or falsify, some of such proposals.





\section*{Acknowledgments} 

F. A. Brito and M. B. Cruz acknowledge Brazilian National Research Council by the financial support.
C. A. S. Silva acknowledges F. A. Brito by his hospitality at Campina Grande Federal University. The authors acknowledge the anonymous referee for the comments and suggestions


\begin{thebibliography}{}


  


  

  
  

  
  



  
	
	\bibitem{Schutz:1985zz}
  B.~F.~Schutz and C.~M.~Will,
 ``Black Hole Normal Modes: A Semianalytic Approach,''
  Astrophys.\ J.\  291 (1985) L33.

\bibitem{s.iyer-prd35}{S. Iyer and C. M. Will, Phys. Rev. D 35, 3621 (1987).}

\bibitem{Mathur:2005zp}
  S.~D.~Mathur,
  ``The Fuzzball proposal for black holes: An Elementary review,''
  Fortsch.\ Phys.\  {\bf 53} (2005) 793
  doi:10.1002/prop.200410203
  [hep-th/0502050].
	
	\bibitem{Nozari:2008gp} 
  K.~Nozari and S.~Hamid Mehdipour,
  ``Quantum Gravity and Recovery of Information in Black Hole Evaporation,''
  Europhys.\ Lett.\  {\bf 84} (2008) 20008
  [gr-qc/0804.4221].

\bibitem{Silva:2008kh}
  C.~A.~S.~Silva,
  ``Fuzzy spaces topology change as a possible solution to the black hole information loss paradox,''
  Phys.\ Lett.\ B {\bf 677} (2009) 318
  doi:10.1016/j.physletb.2009.05.031
  [gr-qc/0812.3171].
	
	


\bibitem{Silva:2010ir}
  C.~A.~S.~Silva and R.~R.~Landim,
  ``A note on black hole entropy, area spectrum, and evaporation,''
  Europhys.\ Lett.\  {\bf 96} (2011) 10007
  doi:10.1209/0295-5075/96/10007
  [gr-qc/1003.3679].
	
	\bibitem{Fazeli:2010zz}
  R.~Fazeli, S.H.~Mehdipour and S.~Sayyadzad,
  ``Generalized Uncertainty Principle in Hawking Radiation of Non-Commutative Schwarzschild Black Ho le,''
  Acta Phys.\ Polon.\ B {\bf 41} (2010) 2365.
	
	\bibitem{Kim:2011fh}
  H.~Kim,
  ``Hawking radiation as tunneling from charged black holes in 0A string theory,''
  Phys.\ Lett.\ B {\bf 703} (2011) 94
  doi:10.1016/j.physletb.2011.07.053
  [hep-th/1103.3133].
	
	\bibitem{Silva:2014jda}
  C.~A.~S.~Silva and R.~R.~Landim,
  ``Fuzzy spaces topology change and BH thermodynamics,''
  J.\ Phys.\ Conf.\ Ser.\  {\bf 490} (2014) 012012.
  doi:10.1088/1742-6596/490/1/012012
	
\bibitem{Modesto:2008im}
  L.~Modesto,
  ``Semiclassical loop quantum black hole,''
  Int.\ J.\ Theor.\ Phys.\  {\bf 49} (2010) 1649
  doi:10.1007/s10773-010-0346-x


  
  \bibitem{Modesto:2009ve}
  L.~Modesto and I.~Premont-Schwarz,
``Self-dual Black Holes in LQG: Theory and Phenomenology'',
  Phys.\ Rev.\ D 80 (2009) 064041.
	
	
	\bibitem{Aragao:2016vuy}
  R.~G.~L.~Arag\~{a}o, C.~A.~S.~Silva,
``Entropic corrected Newton's law of gravitation and the loop quantum black hole gravitational atom'',
  Gen.\ Rel.\ Grav. 48 (2016) no 7, 83.
	arxiv: 1601.04993 [gr-qc]
  
  
  \bibitem{Silva:2015qna}
  C.~A.~S.~Silva,
  Eur.\ Phys.\ J.\ C {\bf 78} (2018) no.5,  409
  doi:10.1140/epjc/s10052-018-5882-1
  [arXiv:1503.00559 [gr-qc]].
  
\bibitem{Regge:1957td}
  T.~Regge and J.~A.~Wheeler,
  Phys.\ Rev.\  {\bf 108} (1957) 1063.
  doi:10.1103/PhysRev.108.1063
	
	
\bibitem{Zerilli:1971wd}
  F.~J.~Zerilli,
 ``Gravitational field of a particle falling in a schwarzschild geometry analyzed in tensor harmonics,''
  Phys.\ Rev.\ D 2 (1970) 2141.

\bibitem{Zerilli:1974ai} 
  F.~J.~Zerilli,
  ``Perturbation analysis for gravitational and electromagnetic radiation in a reissner-nordstrom geometry,''
  Phys.\ Rev.\ D 9 (1974) 860.
  doi:10.1103/PhysRevD.9.860

  
  \bibitem{Moncrief:1974ng}
  V.~Moncrief,
  ``Stability of Reissner-Nordstrom black holes,''
  Phys.\ Rev.\ D 10 (1974) 1057.
 doi:10.1103/PhysRevD.10.1057
  
  \bibitem{Moncrief:1974gw}
  V.~Moncrief,
  ``Odd-parity stability of a Reissner-Nordstrom black hole,''
  Phys.\ Rev.\ D 9 (1974) 2707.
   doi:10.1103/PhysRevD.9.2707
  
  \bibitem{Teukolsky:1972my}
  S.~A.~Teukolsky,
  ``Rotating black holes - separable wave equations for gravitational and electromagnetic perturbations,''
  Phys.\ Rev.\ Lett.\  29 (1972) 1114.
   doi:10.1103/PhysRevLett.29.1114
  
 \bibitem{Teukolsky:1974yv}
  S.~A.~Teukolsky and W.~H.~Press,
  ``Perturbations of a rotating black hole. III - Interaction of the hole with gravitational and electromagnet ic radiation,''
  Astrophys.\ J.\  193 (1974) 443.
   doi:10.1086/153180

\bibitem{s.chandrasekhar-mtbh}{S. Chandrasekhar, The Mathematical Theory of Black Holes, Oxford University, New York (1983).}

\bibitem{Berti:2004md}
  E.~Berti,
  ``Black hole quasinormal modes: Hints of quantum gravity?,''
  Conf.\ Proc.\ C {\bf 0405132} (2004) 145
  [gr-qc/0411025].

\bibitem{Abbott:2016blz}
  B.~P.~Abbott {\it et al.} [LIGO Scientific and Virgo Collaborations],
  ``Observation of Gravitational Waves from a Binary Black Hole Merger,''
  Phys.\ Rev.\ Lett.\  116 (2016) no.6,  061102
  doi:10.1103/PhysRevLett.116.061102
  [arXiv:1602.03837 [gr-qc]].
  
  \bibitem{TheLIGOScientific:2016htt}
  B.~P.~Abbott {\it et al.} [LIGO Scientific and Virgo Collaborations],
  ``Astrophysical Implications of the Binary Black-Hole Merger GW150914,''
  Astrophys.\ J.\  818 (2016) no.2,  L22
  doi:10.3847/2041-8205/818/2/L22
  [arXiv:1602.03846 [astro-ph.HE]].
	
	\bibitem{Abbott:2016nmj}
  B.~P.~Abbott {\it et al.} [LIGO Scientific and Virgo Collaborations],
  ``GW151226: Observation of Gravitational Waves from a 22-Solar-Mass Binary Black Hole Coalescence,''
  Phys.\ Rev.\ Lett.\  116 (2016) no.24,  241103
  doi:10.1103/PhysRevLett.116.241103
  [arXiv:1606.04855 [gr-qc]].
	
	\bibitem{Abbott:2017vtc}
  B.~P.~Abbott {\it et al.} [LIGO Scientific and VIRGO Collaborations],
  ``GW170104: Observation of a 50-Solar-Mass Binary Black Hole Coalescence at Redshift 0.2,''
  Phys.\ Rev.\ Lett.\  118 (2017) no.22,  221101
  doi:10.1103/PhysRevLett.118.221101
  [arXiv:1706.01812 [gr-qc]].
	
	\bibitem{TheLIGOScientific:2017qsa}
  B.~P.~Abbott {\it et al.} [LIGO Scientific and Virgo Collaborations],
  ``GW170817: Observation of Gravitational Waves from a Binary Neutron Star Inspiral,''
  Phys.\ Rev.\ Lett.\  {\bf 119} (2017) no.16,  161101
  doi:10.1103/PhysRevLett.119.161101
  [arXiv:1710.05832 [gr-qc]].
	
	 \bibitem{Konoplya:2011qq}
R.~A.~Konoplya and A.~Zhidenko,
"`Quasinormal modes of black holes: From astrophysics to string theory,"'
Rev.\ Mod.\ Phys.\ 83 (2011) 793
[arXiv:1102.4014 [qr-qc]].
	
	\bibitem{Chirenti:2017mwe}
  C.~Chirenti,
  ``Black hole quasinormal modes in the era of LIGO,''
  arXiv:1708.04476 [gr-qc].
	
 	
	\bibitem{Chen:2011zzi}
  J.~H.~Chen and Y.~J.~Wang,
  ``Complex frequencies of a massless scalar field in loop quantum black hole spacetime,''
  Chin.\ Phys.\ B 20 (2011) 030401.
	
	
\bibitem{Santos:2015gja}
  V.~Santos, R.~V.~Maluf and C.~A.~S.~Almeida,
  ``Quasinormal frequencies of self-dual black holes,''
  Phys.\ Rev.\ D {\bf 93} (2016) no.8,  084047
  doi:10.1103/PhysRevD.93.084047
  [arXiv:1509.04306 [gr-qc]].
  
  

\bibitem{Rovelli:2004tv}
  C.~Rovelli,
  ``Quantum gravity,''
  Cambridge, UK: Univ. Pr. (2004) 455 p


  \bibitem{Sahu:2015dea}
  S.~Sahu, K.~Lochan and D.~Narasimha,
 ``Gravitational lensing by self-dual black holes in loop quantum gravity,''
  Phys.\ Rev.\ D 91 (2015) 063001
  [arXiv:1502.05619 [gr-qc]].
  
  
\bibitem{Kuchar:1994zk} 
  K.~V.~Kuchar,
  ``Geometrodynamics of Schwarzschild black holes,''
  Phys.\ Rev.\ D 50 (1994) 3961
  doi:10.1103/PhysRevD.50.3961
  [gr-qc/9403003].

\bibitem{Thiemann:1992jj}
  T.~Thiemann and H.~A.~Kastrup,
  ``Canonical quantization of spherically symmetric gravity in Ashtekar's selfdual representation,''
  Nucl.\ Phys.\ B 399 (1993) 211
  doi:10.1016/0550-3213(93)90623-W
  [gr-qc/9310012].

\bibitem{Campiglia:2007pr}
  M.~Campiglia, R.~Gambini and J.~Pullin,
  ``Loop quantization of spherically symmetric midi-superspaces,''
  Class.\ Quant.\ Grav.\  24 (2007) 3649
  doi:10.1088/0264-9381/24/14/007
  [gr-qc/0703135].

\bibitem{Modesto:2004xx}
  L.~Modesto,
  ``Disappearance of black hole singularity in quantum gravity,''
  Phys.\ Rev.\ D 70 (2004) 124009
  doi:10.1103/PhysRevD.70.124009
  [gr-qc/0407097].

\bibitem{Bengtsson:1988hm} 
  I.~Bengtsson,
  ``Note on Ashtekar's Variables in the Spherically Symmetric Case,''
  Class.\ Quant.\ Grav.\  5 (1988) L139.
  doi:10.1088/0264-9381/5/10/002

\bibitem{Bojowald:1999eh}
  M.~Bojowald and H.~A.~Kastrup,
  ``Quantum symmetry reduction for diffeomorphism invariant theories of connections,''
  Class.\ Quant.\ Grav.\  17 (2000) 3009
  doi:10.1088/0264-9381/17/15/311
  [hep-th/9907042].

\bibitem{Bojowald:2004ag}
  M.~Bojowald and R.~Swiderski,
  ``The Volume operator in spherically symmetric quantum geometry,''
  Class.\ Quant.\ Grav.\  21 (2004) 4881
  doi:10.1088/0264-9381/21/21/009
  [gr-qc/0407018].

\bibitem{Bojowald:2004si}
  M.~Bojowald and R.~Swiderski,
  ``Spherically symmetric quantum horizons,''
  Phys.\ Rev.\ D 71 (2005) 081501
  doi:10.1103/PhysRevD.71.081501
  [gr-qc/0410147].

\bibitem{Bojowald:2005cb}
  M.~Bojowald and R.~Swiderski,
  ``Spherically symmetric quantum geometry: Hamiltonian constraint,''
  Class.\ Quant.\ Grav.\  23 (2006) 2129
  doi:10.1088/0264-9381/23/6/015
  [gr-qc/0511108].

\bibitem{Gambini:2013ooa}
  R.~Gambini and J.~Pullin,
  ``Loop quantization of the Schwarzschild black hole,''
  Phys.\ Rev.\ Lett.\  110 (2013) no.21,  211301
  doi:10.1103/PhysRevLett.110.211301
  [arXiv:1302.5265 [gr-qc]].


\bibitem{s.hossenfelder-prd81}  

S.~Hossenfelder, L.~Modesto and I.~Premont-Schwarz,
  ``A Model for non-singular black hole collapse and evaporation'',
  Phys.\ Rev.\ D 81 (2010) 044036
  [arXiv:0912.1823 [gr-qc]].

 

\bibitem{Alesci:2011wn}
  E.~Alesci and L.~Modesto,
  ``Particle Creation by Loop Black Holes,''
  Gen.\ Rel.\ Grav.\  46 (2014) 1656
  [arXiv:1101.5792 [gr-qc]].
  
    \bibitem{Carr:2011pr}
  B.~Carr, L.~Modesto and I.~Premont-Schwarz,
  ``Generalized Uncertainty Principle and Self-dual Black Holes'',
  arXiv:1107.0708 [gr-qc].


\bibitem{s.hossenfelder-2012ca} 
   S.~Hossenfelder, L.~Modesto and I.~Premont-Schwarz,
  ``Emission spectra of self-dual black holes'',
  arXiv:1202.0412 [gr-qc].
  
  \bibitem{Silva:2012mt}
  C.~A.~S.~Silva and F.~A.~Brito,
  ``Quantum tunneling radiation from self-dual black holes,''
  Phys.\ Lett.\ B 725 (2013) 45,  456
  [arXiv:1210.4472 [physics.gen-ph]].
  
\bibitem{Anacleto:2015mma}
  M.~A.~Anacleto, F.~A.~Brito and E.~Passos,
 ``Quantum-corrected self-dual black hole entropy in tunneling formalism with GUP,''
  Phys.\ Lett.\ B 749 (2015) 181
   doi:10.1016/j.physletb.2015.07.072
  [arXiv:1504.06295 [hep-th]].
	
\bibitem{Rezzolla:2003ua}
  L.~Rezzolla,
  ``Gravitational waves from perturbed black holes and relativistic stars,''
  ICTP Lect.\ Notes Ser.\  {\bf 14} (2003) 255
  [gr-qc/0302025].
	
\bibitem{Frolov:1998wf} 
  V.~P.~Frolov and I.~D.~Novikov,   in Black hole physics: basic concepts and new developments. Fundamental theories of physics, vol. 96, Kluwer Academic Publishers, 1998
	
\bibitem{hj.blome-pla1984}
  H.~J. Blome and B.~Mashhoon, 
	``The quasi-normal oscillations of a Schwarzschild black hole,''
	Phys. Lett. A. 100, 231 (1984)
  
	


\bibitem{Chandrasekhar:1975zza}
S.~Chandrasekhar and S.~L.~Detweiler,
"`The quasi-normal modes of the Schwarzschild black hole,"'
Proc. \ Roy.\ Soc. \ Lond.\ A 344 (1975) 441.

\bibitem{Gundlach:1993tp}
C.~Gundlach, R.~H.~Price and J.~Pullin,
"`Late time behavior of stellar collapse and explosions: 1. Linearized perturbations,"'
Phys.\ Rev.\ D 49 (1994) 883
[gr-qc/9307009].


\bibitem{Leaver:1985ax}
E.~W.~Leaver,
"`An Analytic representation for the quasi normal mode of Kerr black hole,"'
Proc. \ Roy.\ Soc. \ Lond.\ A 402 (1985) 285.


\bibitem{Leaver:1990zz}
E.~W.~Leaver,
"`Quasinormal modes of Reissner-Nordestrom black holes,"'
Phys.\ Rev.\ D 41 (1990) 2986.


\bibitem{Nollert:1993zz}
H.~P.~Nollert,
"`Quasinormal modes of Schwarzschild black holes: The determination of quasinormal frequencies with very large imaginary parts,"'
Phys.\ Rev.\ D 47 (1993) 5253.

\bibitem{Berti:2009kk}
E.~Berti, V.~Cardoso and A.~O.~Starinets,
"`Quasinormal modes of black holes and black branes,"'
Class.\ Quant. \ Grav.\ 26 (2009) 163001
[arXiv:0905.2975 [gr-qc]].

\bibitem{Kokkotas:1999bd}
K.~D.~Kokkotas and B.~G.~Schimidt,
"`Quasinormal modes of stars and black holes,"'
Living Rev.\ Rel,\ 2 (1999) 2
[ge-qc/9909058]

\bibitem{Nollert:1999ji}
H.~P.~Nollert,
"`TOPICAL REVIEW: Quasinormal modes: the characteristic 'sound' of black holes and neutrom stars,"'
Class.\ Quant. \ Grav.\ 16 (1999) R159.

	
\bibitem{Konoplya:2004ip} 
  R.~A.~Konoplya,
  ``Quasinormal modes of the Schwarzschild black hole and higher order WKB approach,''
  J.\ Phys.\ Stud.\  {\bf 8}, 93 (2004).
	
	\bibitem{Perez:2017cmj}
  A.~Perez,
  Rept.\ Prog.\ Phys.\  {\bf 80} (2017) no.12,  126901
  doi:10.1088/1361-6633/aa7e14
  [arXiv:1703.09149 [gr-qc]].
	
	\bibitem{Ashtekar:2018lag}
  A.~Ashtekar, J.~Olmedo and P.~Singh,
  Phys.\ Rev.\ Lett.\  {\bf 121} (2018) no.24,  241301
  doi:10.1103/PhysRevLett.121.241301
  [arXiv:1806.00648 [gr-qc]].
	
	\bibitem{Ashtekar:2018cay}
  A.~Ashtekar, J.~Olmedo and P.~Singh,
  Phys.\ Rev.\ D {\bf 98} (2018) no.12,  126003
  doi:10.1103/PhysRevD.98.126003
  [arXiv:1806.02406 [gr-qc]].
	

	
	

\end{thebibliography}
\end{document}